\documentclass[12pt,sort&compress]{elsarticle}
\pdfoutput=1
\usepackage{amsmath}
\usepackage{graphicx}
\usepackage{bm}
\usepackage{epsfig}
\usepackage{psfrag}
\usepackage{multirow}
\usepackage{dcolumn}
\usepackage{amssymb}
\usepackage{subfig}
\usepackage{hyperref}
\usepackage{xfrac}
\usepackage{scalefnt}

\makeatletter
\def\ps@pprintTitle{%
 \let\@oddhead\@empty
 \let\@evenhead\@empty
 \def\@oddfoot{}%
 \let\@evenfoot\@oddfoot}
\makeatother

\journal{Nuclear Instruments and Methods in Physics Research}

\newcommand{\chisq}  {{\ensuremath{\chi^2}}}
\newcommand{\Nsigma}   {{\ensuremath{N_\sigma}}}

\newcommand {\pbar}	{{\ensuremath{\bar p}}}
\newcommand {\ppbar}	{{\ensuremath{p\pbar}}}
\newcommand {\tbar}     {{\ensuremath{\bar t}}}
\newcommand {\ttbar}    {{\ensuremath{t\tbar}}}

\newcommand {\qbar}     {{\ensuremath{\bar q}}}
\newcommand {\qqbar}    {{\ensuremath{q\qbar}}}
\newcommand {\itW}      {\ensuremath{W}}
\newcommand {\Whboson}  {\itW-boson}
\newcommand {\Wboson}  {\itW\ boson}
\newcommand {\pjet}    {{\textrm{+jets}}}
\newcommand {\lpj}      {{\ensuremath{l\pjet}}}
\newcommand {\lptj}      {{\ensuremath{l\textrm{+}3\textrm{\,jets}}}}
\newcommand {\lpgefj}      {{\ensuremath{l}+\ensuremath{\geq}4\,jets}}

\newcommand {\GeV}        {{\ensuremath{\,\textrm {GeV}}}}
\newcommand {\TeV}        {{\ensuremath{\,\textrm {TeV}}}}

\newcommand {\pt}         {{\ensuremath{p_T}}}
\newcommand {\dy}         {{\ensuremath{\Delta y}}}
\newcommand {\qt}         {{\ensuremath{\vec{q_T}}}}
\newcommand {\qz}         {{\ensuremath{q_z}}}
\newcommand{\emiss} {{/\!\!\!\!E}}
\newcommand{\met}   {{\ensuremath{\emiss_T}}}
\newcommand{\mttbar} {{\ensuremath{m_\ttbar}}}
\newcommand{\gmtt}   {{\ensuremath{m_{t\bar{t}}^\text{gen}}}}
\newcommand{\amtt}   {\mttbar} 
\newcommand{\mpeak}   {{\ensuremath{m_{t\bar{t}}^\text{peak}}}}
\newcommand{\mproxy}   {{\ensuremath{m_p}}}
\newcommand{\ahat}   {{\ensuremath{\hat{\alpha}}}}
\newcommand {\mW}      {{\ensuremath{M_W}}}

\newcommand \smaller[2][0.72]{{\scalefont{#1}#2}} 
 \newcommand {\DZ}     {{D0}} 
 \newcommand {\this}     {{paper}} 
 \newcommand {\mcatnlo}  {\smaller{MC@NLO}}
 \newcommand {\ie}       {{\textrm{i.e.}}}
 \newcommand {\eg}       {{\textrm{e.g.}}}
 
 \newcommand {\etal}     {{\emph{et al.}}}

\newcommand {\tablestrut} {\rule{0pt}{2.5ex}}
\newcommand{\centercell}[1]{\multicolumn{1}{c}{#1}}
\newcommand{\head}[1]{\centercell{\bfseries\boldmath#1}}
\newcommand{\multihead}[2]{\multicolumn{#1}{c}{\bfseries\boldmath#2}}

\newcommand{\dd}{\; \mathrm{d}} 
\DeclareMathOperator\erf{erf}

\begin{document}

\begin{frontmatter}

\title{Reconstructing \ttbar\ events with one lost jet}

\author{Regina Demina}
\ead{regina@pas.rochester.edu}
\author{Amnon Harel, Douglas Orbaker}

\address{Department of Physics and Astronomy, University of Rochester, Rochester NY, United States of America}

%
%
%
\begin{abstract}
We present a technique for reconstructing the kinematics of pair-produced top quarks that decay
to a charged lepton, a neutrino and four final state quarks in the subset of events where only
three jets are reconstructed.
We present a figure of merit that allows for a fair comparison of reconstruction algorithms without
requiring their calibration.
The new reconstruction of events with only three jets is fully competitive with the 
full reconstruction typically used for four-jet events.
\end{abstract}

\begin{keyword}
top \sep reconstruction \sep partial
\end{keyword}

\end{frontmatter}


%
%
\section{Introduction}
\label{sec:intro}
Several problems in top quark physics require a full reconstruction of the kinematics of the
top quark--antiquark pair. For example, to measure the forward-backward (or charge) asymmetry in \ttbar\ 
production, it is essential to know the direction of both the top quark and the antiquark.

We consider \ttbar\ events where each top quark decays into a $b$ quark and a \Wboson,
and where one \Wboson\ decays hadronically ($W\to q'\bar{q}$) and one \Wboson\ decays leptonically ($W\to l\nu$).
We classify top quarks as ``leptonic'' or ``hadronic'', based on the mode of the \Whboson\ decay. 
The final state contains a lepton, a neutrino and four quarks that subsequently shower and 
hadronize into jets. 
This channel is commonly referred to as ``\lpj''. 

The four final state quarks do not always yield four reconstructed jets, which is the case, for example,
when one of the quarks is too soft or when the angular separation between two of them is small. 
Though the signal purity is lower in the sample of events with exactly three jets than in the sample
with at least four jets, it is still useful for measuring top properties~\cite{bib:afbl,bib:afb}, effectively
increasing the sample size by $\approx55\%$.
Furthermore, extending the event selection of a top property measurement in the \lpj\ channel
to include three-jet events can reduce the acceptance bias~\cite{bib:afbl} and reduces systematic
uncertainties related to jet reconstruction, as events with one unreconstructed jet are still used.
The three-jet sample can also be interesting in its own right. For example, at the Large Hadron
Collider (LHC) \ttbar\ production of events with at least four jets is dominated by initial states
that contain gluons, while the three-jet sample is enriched in \ttbar\ pairs produced from
\qqbar\ initial states.

When one of the jets from top decay is lost, it is not possible to fully reconstruct the \ttbar\ decay chain,
which has so far limited the use of lepton plus three jets (\lptj) events to measurements of observables such as the 
production cross section~\cite{bib:xsec} and the rapidity\footnote{
The rapidity $y$ is defined as $y\left(\theta,\beta\right)=\frac{1}{2}\ln\left[\left(1+\beta\cos\theta\right)\right.$ 
$\left./\left(1-\beta\cos\theta\right)\right]$,
where $\theta$ is the polar angle and $\beta$ is the ratio of a particle's momentum to its energy.
The pseudorapidity $\eta$ is defined as $-\ln\tan\frac{\theta}{2}$. 
In this \this, pseudorapidities are defined relative to the center of the detector, while rapidities and all other
angles are defined with respect to the primary collision point.}
of the lepton~\cite{bib:afbl}.

The kinematics of the \ttbar\ events with a lepton and at least four jets (\lpgefj) is fully reconstructed
by matching four of the jets to the four final state quarks from \ttbar\ decay (for example, see~\cite{bib:hitfit}).
As for events with at least four jets, the main challenge in fully reconstructing three-jet events
is to disentangle the two top-quark decay chains. That is, the main challenge is to 
match the observed jets with the quarks from \ttbar\ decay, though with only three jets
available a perfect 1-to-1 correspondence is impossible and partial matchings are used instead.

In this \this\ we present a method to infer the direction and kinematics of the top quark and antiquark
in \lpj\ events where only three jets are reconstructed, 
and demonstrate the application of the method to simulated \ttbar\ events.
We focus on $\ppbar\to\ttbar$ production at a center of mass energy of $1.96\TeV$, as in the Tevatron. 
About half of the \ttbar\ \lpj\ events produced at the Tevatron contain only three jets. 

The main steps of the method have been described in Ref.~\cite{bib:afb}, where it is used 
to reconstruct the directions of the top quark and antiquark and the invariant mass of the \ttbar\ system.
This \this\ provides the details of the method and quantifies its performance.
We discuss the selection of the events in Section~\ref{sec:sel}.
In Section~\ref{sec:reco3j} we detail the method to partially reconstruct the
\ttbar\ pair using the invariant mass of various combinations of jets and jet lifetime observables~\cite{bib:btags}.
We compare the performance of different reconstruction algorithms in Section~\ref{sec:performance}, 
for which we introduce a new figure of merit (FOM).

The reconstruction of \ttbar\ events at the LHC poses different problems.
The typical jet-selection threshold for the LHC is transverse momentum $\pt>30\GeV$, and only in $40(2)\%$
\footnote{Here and later the presented fractions are representative,
  but are given only as example as they depend on the detector, the event
  selection, the jet algorithm, etc. The uncertainty on the last significant figure is given in parenthesis.}
of the \ttbar\ events do all four jets associated with \ttbar-decay quarks pass this threshold.
Yet, due to the initial state radiation (ISR) only a small fraction of \ttbar\ events produced at the LHC end up in
the three-jet sample. For the LHC \lptj\ events, a method similar to that presented here may suffice.
Roughly $40\%$ of the LHC \lpgefj\ events contain only three jets associated with \ttbar\ decay quarks,
with the other jets due to ISR. An extension of the algorithm discussed in this paper could be used to
partially reconstruct these events and thus increase the number of reconstructed events
by approximately a factor of two.

%
%
\section{Samples and selection}
\label{sec:sel}

The results shown in this paper are based on simulated $\ppbar\to\ttbar$ events
with a collision center-of-mass energy of $\sqrt{s}=1.96\TeV$. 
The events were simulated with the \mcatnlo\
event generator~\cite{bib:mcatnlo} and processed through a detector simulation and
object reconstruction that largely correspond to but are not identical to that of the \DZ\ experiment. 
In particular, some of the 
quality selection criteria are not applied since they are not relevant for the development
of the method.

Simulated energy deposits in the calorimetry are clustered into jets using 
the ``Run II Midpoint cone algorithm''~\cite{bib:jetalgo}
with a cone radius of $0.5$ in the $y$-$\phi$ plane, where $\phi$ is the azimuthal angle
and $y$ is the rapidity.
We select jets with $\pt>20\GeV$ and with pseudorapidity $|\eta|<2.5$. 

We select leptons from electron and muon candidates with
$\pt>20\GeV$ and with $|\eta_e|<1.1$ or $|\eta_\mu|<2.0$.
We then select events with exactly one lepton and exactly three jets.
We require that the 
transverse momentum imbalance measured by the calorimetry, \met,
is greater than $20\GeV$.
We reject events where the \met\ is closely aligned with the 
lepton and events with \met $>500\GeV$.
These two cuts suppress multijet background and events with misreconstructed \met, respectively.

Generally, the signal purity is lower in the sample of events with exactly three jets than in the sample
with at least four jets. However additional selection criteria,
e.g. identification of jets associated with $b$-quarks ($b$-tagging), can improve the situation,
making the \lptj\ sample useful for measuring top properties.
In particular, in~\cite{bib:afbl,bib:afb} it was shown that purity of \lptj\ sample with two $b$-tags is
similar to that of \lpgefj\ with one $b$-tag.

We further categorize the selected events by how well the reconstructed jets
match the quarks from \ttbar\ decay, as that affects the quality of reconstruction.
We consider a jet to be matched to a quark when their angular separation
$\Delta R=\sqrt{(\Delta y)^2+(\Delta\phi)^2}$ is less than 0.5.
We classify  an event as ``matchable''  if all \ttbar\ decay products  assumed  
to be present by the reconstruction algorithm were matched to reconstructed objects. 

For the reconstruction of \lpgefj\ events at the Tevatron~\cite{bib:hitfit}, a matchable event is the one in which
the four jets of highest \pt\ match the four final state quarks from \ttbar\ decay.
Only $55(1)\%$ of the \lpgefj\ events at the Tevatron are matchable.
In the context of this \this\ a \lptj\ event is considered  matchable if one jet matches the $b$ quark from 
of the leptonic top quark decay
and the two other jets match two of the three quarks from the decay of the hadronically decaying top quark.
$20(1)\%$ of the \lptj\ events are classified as unmatchable because the $b$ jet from the leptonic top decay,
which is essential to the described algorithm, is lost.
In 4.0(2)\% of the events two jets were lost, while an extra one was gained from initial or final state radiation.
Thus, $76(1)\%$ of the \lptj\ events are considered matchable.

%
%
\section{\boldmath Reconstructing \ttbar\ in \lptj\ events}
\label{sec:reco3j}

For almost half of the simulated $\ppbar\to\ttbar$ pairs
that decay in the \lpj\ channel, only three jets are reconstructed.
In our study scenario, two quarks yield a single jet due to an accidental overlap
in $\approx18\%$ of these \lptj\ events.  
One of the quarks is too forward (high $|\eta|$) to yield a selected jet
in $\approx8\%$ of the events.
In the remaining $\approx74\%$ of the events, either one of the quarks 
was too soft (low \pt) to yield a selected jet or a jet was lost due to 
reconstruction and identification inefficiencies.

\begin{figure}[htbp]
\begin{center}
\includegraphics[width=0.5\linewidth,viewport=65 55 705 490]{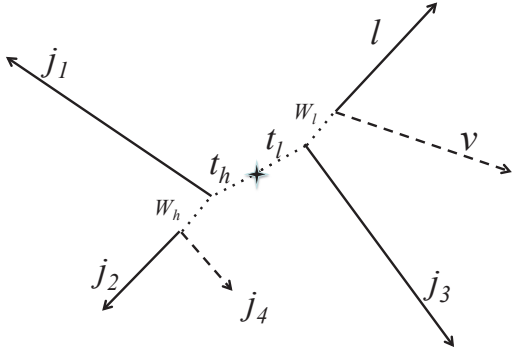}
\end{center}
\caption{
Cartoon depicting an example of \ttbar\ decay: the leptonic top quark ($t_l$) decays to $b$-jet $j_3$ and to a \Wboson\ 
which decays to a lepton and a neutrino; the hadronic top quark ($t_h$) decays to $b$-jet $j_1$ and to a \Wboson\
which decays hadronically to jets $j_2$ and $j_4$. In the depicted \lptj\ event, $j_4$ is lost. }
\label{fig:cartoon}
\end{figure}

In Fig.~\ref{fig:cartoon} we show a schematic of a possible \ttbar\ decay process. 
Instead of trying to infer the kinematics of the missing or merged jet in a \lptj\ event, 
we partially reconstruct the \ttbar\ system by neglecting this jet altogether.
Though there is some experimental sensitivity to the presence of two quarks in a single
jet, e.g., through the jet width and mass, we found it too weak to be useful.
Thus we do not attempt to ``unmerge'' any of the jets and assign two quarks to it.
Events in the \lpgefj\ channel are often reconstructed using a ``kinematic fit'' algorithm, which modifies 
the measured momenta to satisfy the known resonance masses (\eg\ Ref.~\cite{bib:hitfit}).
Given that we neglect the missing jet, such refinements are of little use for \lptj\ events.
Thus we employ a simpler approach to partially reconstruct the \ttbar\ system in \lptj\ events.
%
\subsection{Reconstructing the leptonic \Wboson}
\label{sec:recoWj}
We start by reconstructing the leptonically decaying \Wboson\ using 
the lepton momentum and the \met.
The neutrino momentum in the plane transverse to the beam direction, \qt, 
is initially set equal to the \met.
The longitudinal component of neutrino momentum, \qz, is calculated using a constraint on the \Whboson\ mass, \mW. 
The resultant quadratic equation can have two solutions, which creates a two-fold ambiguity. 
Both solutions are considered. 

Following Ref.~\cite{bib:ttres}, when the discriminant of the quadratic equation for \qz\ is negative, 
we scale \qt\ to satisfy the \mW\ constraint with a discriminant equal to zero.
This results in another quadratic equation which yields two solutions for the scale, at least one of which is positive.
When both solutions are positive, we use the one
that is closer to unity.
%
\subsection{Reconstructing the top-quark candidates}
\label{sec:recotj}
The next step is to form leptonic and hadronic top quark candidates. 
To do so, we assume that the lost jet is from the decay of the hadronic top quark.
One of the jets is combined with the leptonic \Wboson\ to form a leptonic top candidate. 
The two remaining jets are combined to form a ``proxy'' for the hadronic top quark,
which serves instead of a fully reconstructed candidate.
The assignment is completely defined by the choice of leptonic $b$ jet.
If the previous step yielded two \qz\ solutions, for each assignment
we choose the solution where the combination of the leptonic $b$ jet,
the lepton and the neutrino yields an invariant mass closer to the nominal top quark mass~\cite{bib:topmass}. 
%
\subsubsection{$\chi^2$ method}
\label{sec:chi2j}
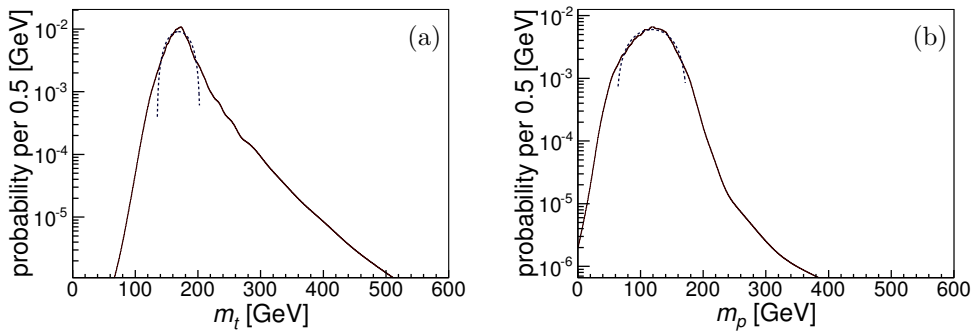
\begin{figure}[htbp]
\begin{center}
\includegraphics[width=0.48\linewidth]{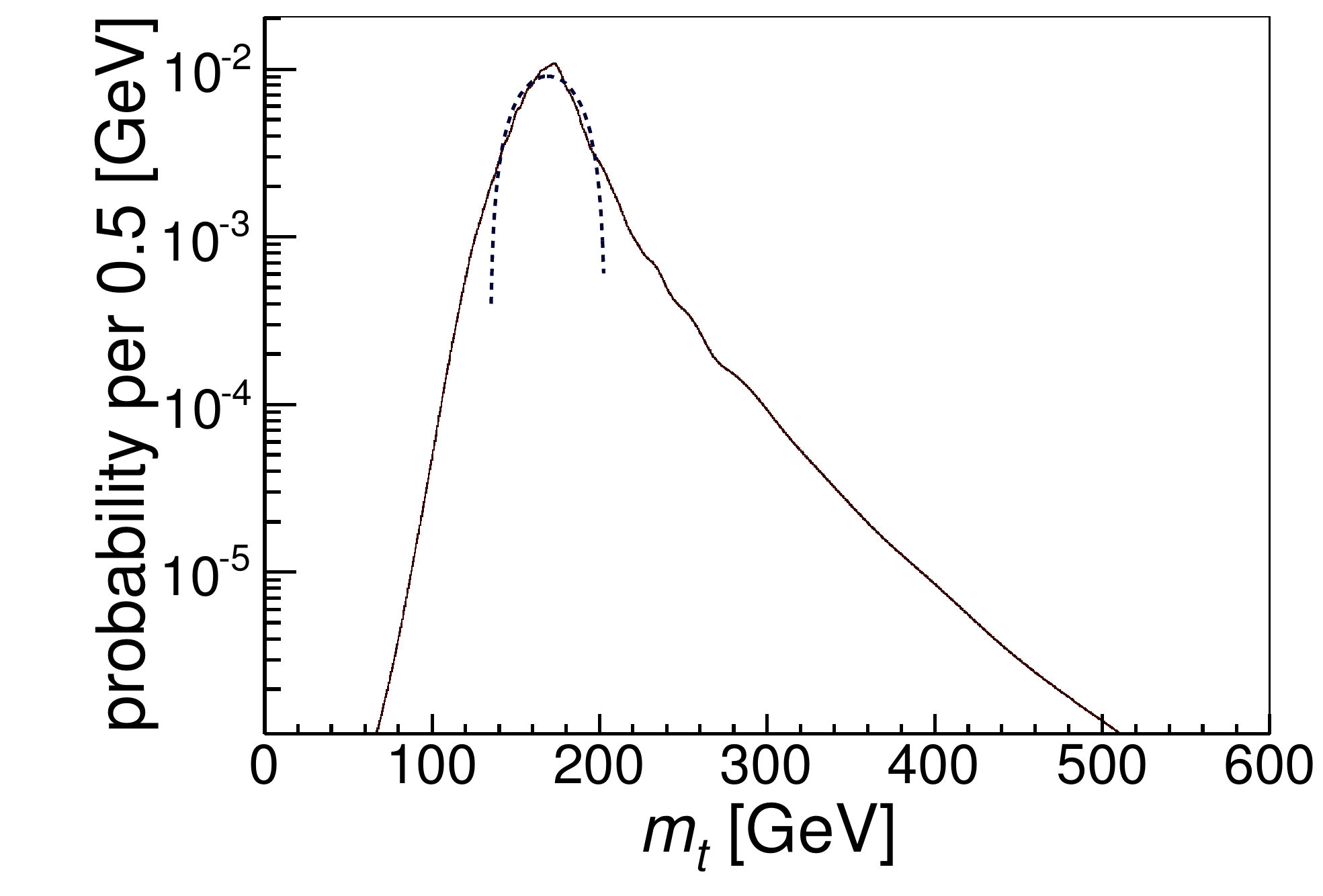}
\includegraphics[width=0.48\linewidth]{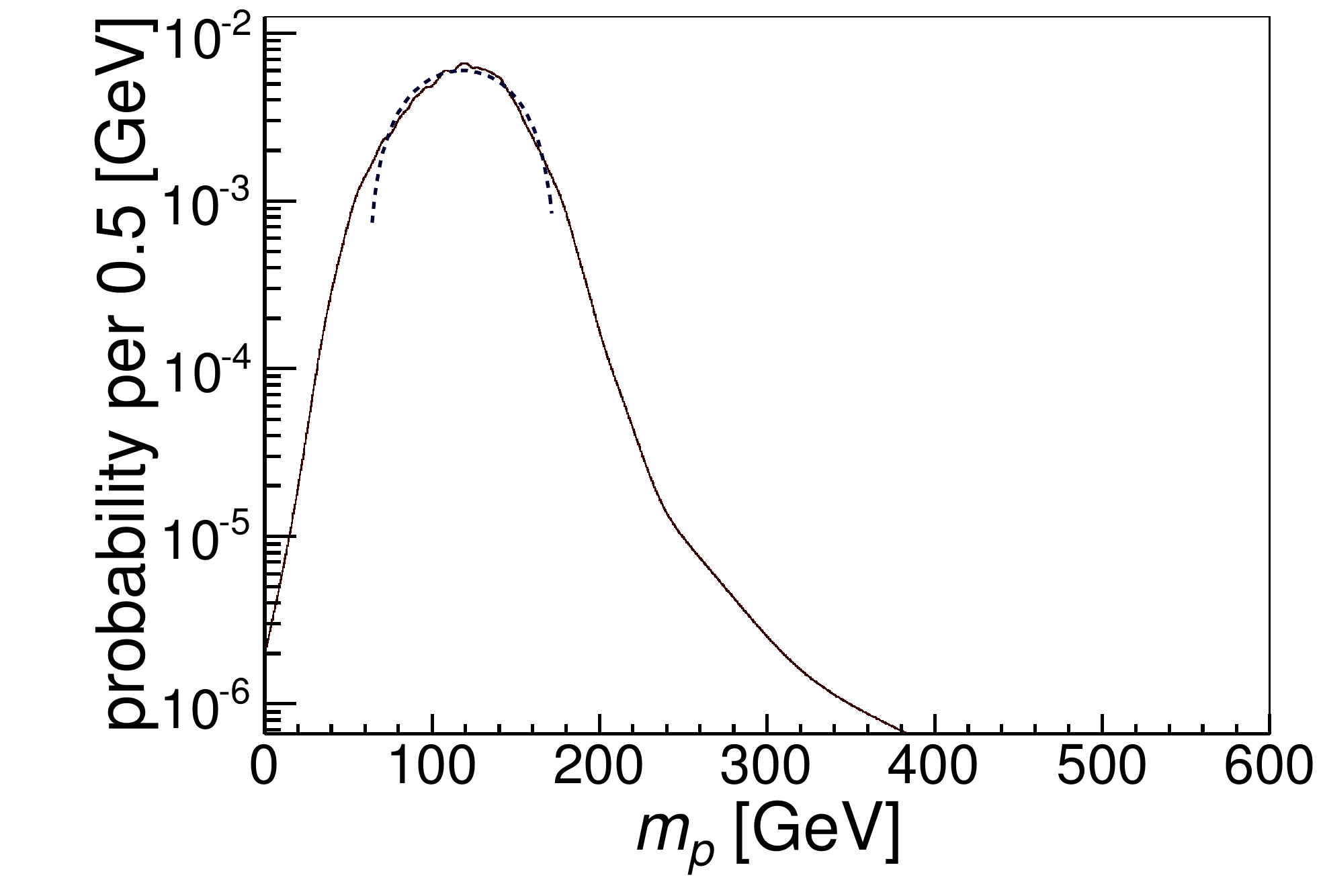}
\end{center}
\begin{picture}(0,0)(0,0)
\put (170,151){\subfloat[][]{\label{subfig:mf_a}}}
\put (361,151){\subfloat[][]{\label{subfig:mf_b}}}
\end{picture}
\vspace{-1.2cm}
\caption{
Invariant mass distribution of (a) lepton, neutrino and $b$ jet from leptonically decaying top quark, 
and (b) invariant mass of two remaining jets from the hadronic top decay. 
In both cases, fits to Gaussian distributions are shown by the dashed curves.
}
\label{fig:mass_fits}
\end{figure}

Invariant mass distributions on both the leptonic and hadronic sides have characteristic 
shapes as shown in Fig.~\ref{fig:mass_fits}. Both can be used to find the best jet assignment.
The distributions were made using an adaptive kernel estimator~\cite{bib:ake}.

A simple way to choose an assignment is to use a \chisq\ test statistic for the masses reconstructed
for the leptonic top candidate ($m_t$) and for the proxy ($m_p$):
\begin{equation}
\chisq=\left(\frac{m_t-m_t^0}{\sigma_t}\right)^2+\left(\frac{m_p-m_p^0}{\sigma_p}\right)^2,
\label{eq:chisq}
\end{equation}
where  $m_t^0$ ($m_p^0$) and $\sigma_t$ ($\sigma_p$) are the mean and width of the Gaussian fits for 
leptonic (proxy) masses shown in Fig.~\ref{fig:mass_fits}.   
This approach picks the correct assignment in $66.0(1)\%$ of the cases where such an assignment exists.
Below we discuss more detailed treatments that improve upon this basic technique.
%
\subsubsection{Complete likelihood method}
\label{sec:lhoodj}
We improve the choice of the assignment by replacing the \chisq\ with a likelihood
function. The likelihood formalism allows us to take into account additional information.
The use of the invariant masses of the incorrect assignments, which too have distinct shapes, 
is detailed below. 
The use of ``$b$-tagging'' observables that attempt to identify jets likely to arise from a 
$b$ quark is detailed further on.

\begin{figure}[htbp]
\begin{center}
\includegraphics[width=0.6\linewidth]{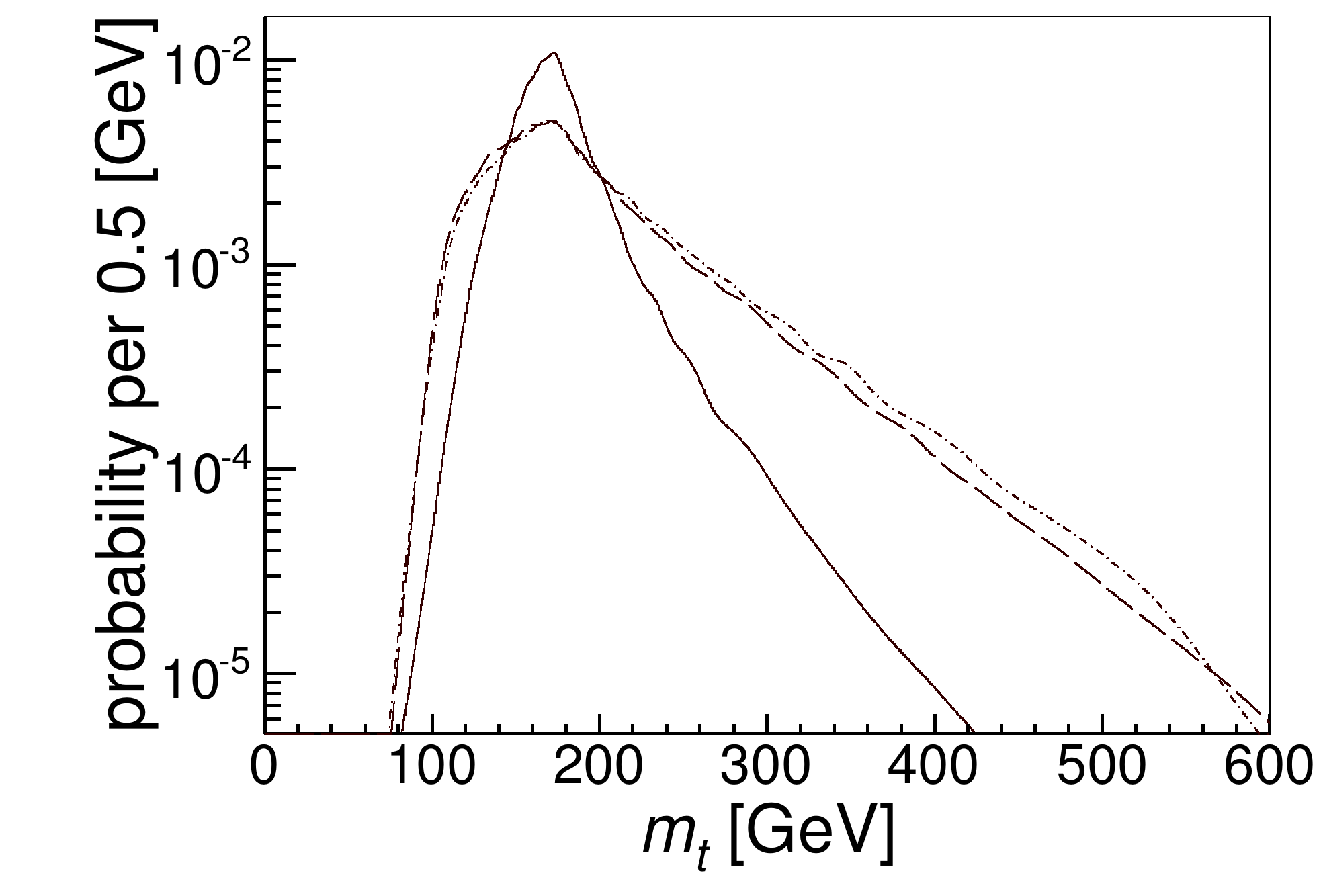}
\end{center}
\vspace{-0.6cm}
\caption{
Distributions of the mass of the leptonic top candidate, 
which comprises the lepton, neutrino and leptonic $b$-jet candidate.
The distribution is shown for events where the jet assigned 
to the leptonic $b$ quark is the correct one ($P_{t:l}$, solid curve),
the hadronic $b$ jet ($P_{t:h}$, dot-dashed curve), 
or a jet from hadronic \Whboson\ decay ($P_{t:q}$, dashed curve).
}
\label{fig:mass_lep}
\end{figure}

Figure~\ref{fig:mass_lep} shows the distributions in top candidate mass on the leptonic side for three situations: 
when the leptonic \Wboson\ is (correctly) combined with the $b$ jet from leptonic top decay ($P_{t:l}$),
when it is (wrongly) combined with the hadronic $b$ jet ($P_{t:h}$), and when it is 
(wrongly) combined with a jet from hadronic \Whboson\ decay ($P_{t:q}$).   
Using the distinct shape of a presumably  ``incorrect'' 
assignment means we need to keep track of two types of assignments which may disagree. We will introduce
notation for the assignment used to combine the jets into the mass observables 
and for the assignment hypothesized to be correct.

\begin{figure}[htbp]
\begin{center}
\includegraphics[width=0.6\linewidth]{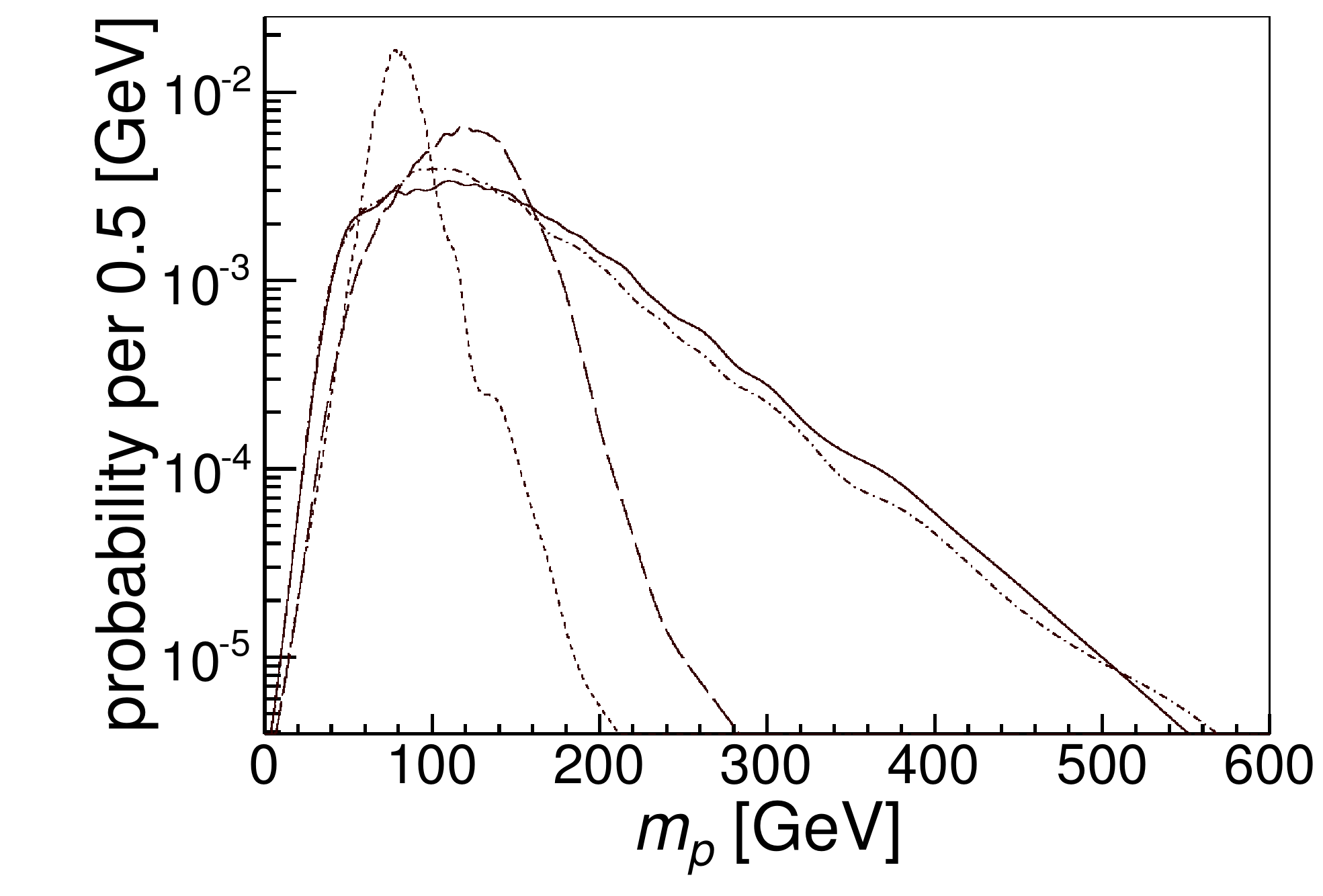}
\end{center}
\vspace{-0.6cm}
\caption{
Distributions of the mass of the proxy for the hadronic top quark, 
which comprises two jets.
The distribution is shown for events where the jets assigned to the proxy
are the hadronic and leptonic $b$ jets ($P_{p:hl}$, solid curve), 
the leptonic $b$ jet and a jet from \Whboson\ decay ($P_{p:lq}$, dot-dashed curve), 
the hadronic $b$ jet and a jet from \Whboson\ decay ($P_{p:hq}$, long dashes),
or both jets from the \Whboson\ decay ($P_{p:qq}$, short dashes).
In the last case, the \itW\ resonance is clearly seen.
}
\label{fig:mass_had}
\end{figure}

Depending on which jet is lost and which jet is picked to form the leptonic top candidate there are 
four possible two-jet combinations for the proxy side. The probability distributions for the invariant mass on the proxy side are shown in Fig.~\ref{fig:mass_had} for
hadronic and leptonic $b$ jets ($P_{p:hl}$), 
leptonic $b$ jet and a jet from \Whboson\ decay($P_{p:lq}$), 
hadronic $b$ jet and a jet from \Whboson\ decay($P_{p:hq}$), and both jets from \Whboson\ decay($P_{p:qq}$).
The first two combinations are incorrect, as they include the leptonic $b$ jet.
The last two combinations are correct, and under the assumption that the 
leptonic $b$ jet was reconstructed, they cannot both be available in the same event.

These shapes can be used to maximize the probability $P$ of selecting the correct assignment 
$a$ given the data $d$, which according to Bayes' theorem is: 
\begin{equation}
P\left(a\mid d\right) 
= \frac{P\left(d\mid a\right)P\left(a\right)}{\sum\limits_b P\left(d\mid b\right)P\left(b\right)}
= \frac{P\left(d\mid a\right)}{\sum\limits_b P\left(d\mid b\right)},
\label{eq:prob}
\end{equation}
where $b$ is any assignment and the second equality uses the fact that a priori all assignments
are equally probable.

There are three possible jet assignments per event ($i=1,2,3$), 
corresponding to the choice of the candidate for the leptonic $b$ jet. 
Each event is characterized by three possible masses on the leptonic side ($t_1,t_2,t_3$) and 
three possible masses on the proxy side ($p_1,p_2,p_3$). 
In addition to this kinematic information, 
$b$-tagging algorithms~\cite{bib:btags} can also help to identify the origins of the jets.
The results of the $b$-tagging algorithms can usually be expressed as
a single continuous variable per jet, 
which discriminates between light and $b$-flavored jets.
We label the $b$-tagging discriminant for the $i$-th jet as $b_i$. 
Thus, data are presented by nine variables: 
\begin{equation}
d=(t_1,t_2,t_3;p_1,p_2,p_3;b_1,b_2,b_3)
\end{equation}

In matchable events the lost jet is either the hadronic $b$ jet or a jet from hadronic \Whboson\ decay. 
We label the former as $Q=b_l qq$ and  the latter $H=b_lb_h q$. 
For a matchable event, 
the probability for assignment $a$ is a weighted sum of the probabilities of $H$ and $Q$ types:
\begin{equation}
P\left(d\mid a\right)=(1-f_Q) P\left(d\mid a,H\right)+f_Q P\left(d\mid a,Q\right),
\end{equation}
where $f_Q$ is the fraction of matchable events that are type $Q$, which 
in our study scenario is $20.5(2)\%$.

Each jet assignment hypothesis specifies the type of each jet: 
either a $b$ jet, or a jet from hadronic \Whboson\ decay. 
The latter category includes jets that arise from $c$ quarks, and are somewhat
similar to $b$ jets~\cite{bib:btags}. 
The correlations between the $b$-tagging discriminants ($b_j$) are small.
Furthermore, these correlations are mostly independent of the true jet flavors, hence
they are irrelevant for our purposes.
Thus, the $b$-tagging probabilities can be factorized:
\begin{align}
  P\left(d\mid a, C\right) &= P\left(t_1,t_2,t_3;p_1,p_2,p_3\mid a, C\right)P\left(b_1,b_2,b_3\mid a, C\right)\\
                           &= P\left(t_1,t_2,t_3;p_1,p_2,p_3\mid a, C\right)\prod_{j=1}^3 P\left(b_j\mid a,C\right)
\label{eq:b_factorized}
\end{align}
where $C=H$ or $Q$ is the hypothesized class of the event.
By neglecting the correlations between the remaining variables we can factorize the first two terms into
six of the one-dimensional distributions shown in Figs.~\ref{fig:mass_lep} and~\ref{fig:mass_had}
($P_{t:y}$ and $P_{p:y}$):
\begin{equation}
   P\left(d\mid a, C\right) = \prod_{j=1}^3 P_{t:f\left(j,a,C\right)}\prod_{j=1}^3 P_{p:g\left(j,a,C\right)}\prod_{j=1}^3 P\left(b_j\mid a,C\right)
\label{eq:factorized}
\end{equation}
where $f\left(j,a,C\right)\in\left\{l,h,q\right\}$ gives the type of the $j$-th jet 
(\ie, the jet assumed to be the leptonic $b$ jet when building the $t_j$ observable) according to assignment
$a$ and event class $C$, 
and $g\left(j,a,C\right)\in\left\{hq,lq,hl,qq\right\}$ gives the types of the non-$j$-th jets 
(\ie, the jets combined to form the proxy for the $p_j$ observable) according to $a$ and $C$.
Though we neglected some of the correlations between the observables in Eq.~\ref{eq:factorized}, 
the structure of the likelihood preserves
the dominant correlations, such as having at most one \Whboson\ resonance, and the correlation
between the presence of a \Whboson\ resonance and the $b$-tagging variables.
Using the described algorithm,
the correct jet assignment is chosen for $69.1(2)\%$ of the matchable events, which is to be compared to $66.0(1)\%$ of 
correct assignments using a simple $\chi^2$ method discussed in Section~\ref{sec:chi2j}.

Returning to the example of Fig.~\ref{fig:cartoon}, the following terms help identify the correct 
event class ($H$) and assignment ($a=3$, i.e. $j_3$ is the leptonic $b$ jet):
 \begin{itemize}
 \item  the invariant mass formed by combining the leptonic \itW\ candidate ($W_l$) and the jet $j_1$, $t_1=m(W_l+j_1)$,
   should be consistent with the $P_{t:h}$ distribution from Fig.~\ref{fig:mass_lep};
 \item  $t_2=m(W_l+j_2)$ should be consistent with $P_{t:q}$ (same figure);
 \item $t_3=m(W_l+j_3)$ should be consistent with $P_{t:l}$ (same figure);
 \item the invariant mass formed by the jets $j_2$ and $j_3$, $p_1=m(j_2+j_3)$, should be consistent 
   with the $P_{p:lq}$ distribution from Fig.~\ref{fig:mass_had};
 \item $p_2=m(j_1+j_3)$, invariant mass of leptonic $b$ jet and a light jet should be consistent with $P_{p:hl}$ (same figure);
 \item $p_3=m(j_1+j_2)$, invariant mass of leptonic and hadronic $b$ jets should be consistent with $P_{p:hq}$ (same figure);
 \item $b_1$, the $b$-tagging discriminant of $j_1$, should be consistent with the distribution for a $b$ jet;
 \item $b_2$  should be consistent with the distribution for a jet from hadronic \Whboson\ decay;
 \item $b_3$  should be consistent with the distribution for a $b$ jet.
 \end{itemize}

\begin{figure}[htbp]
\begin{center}
\includegraphics[width=0.48\linewidth]{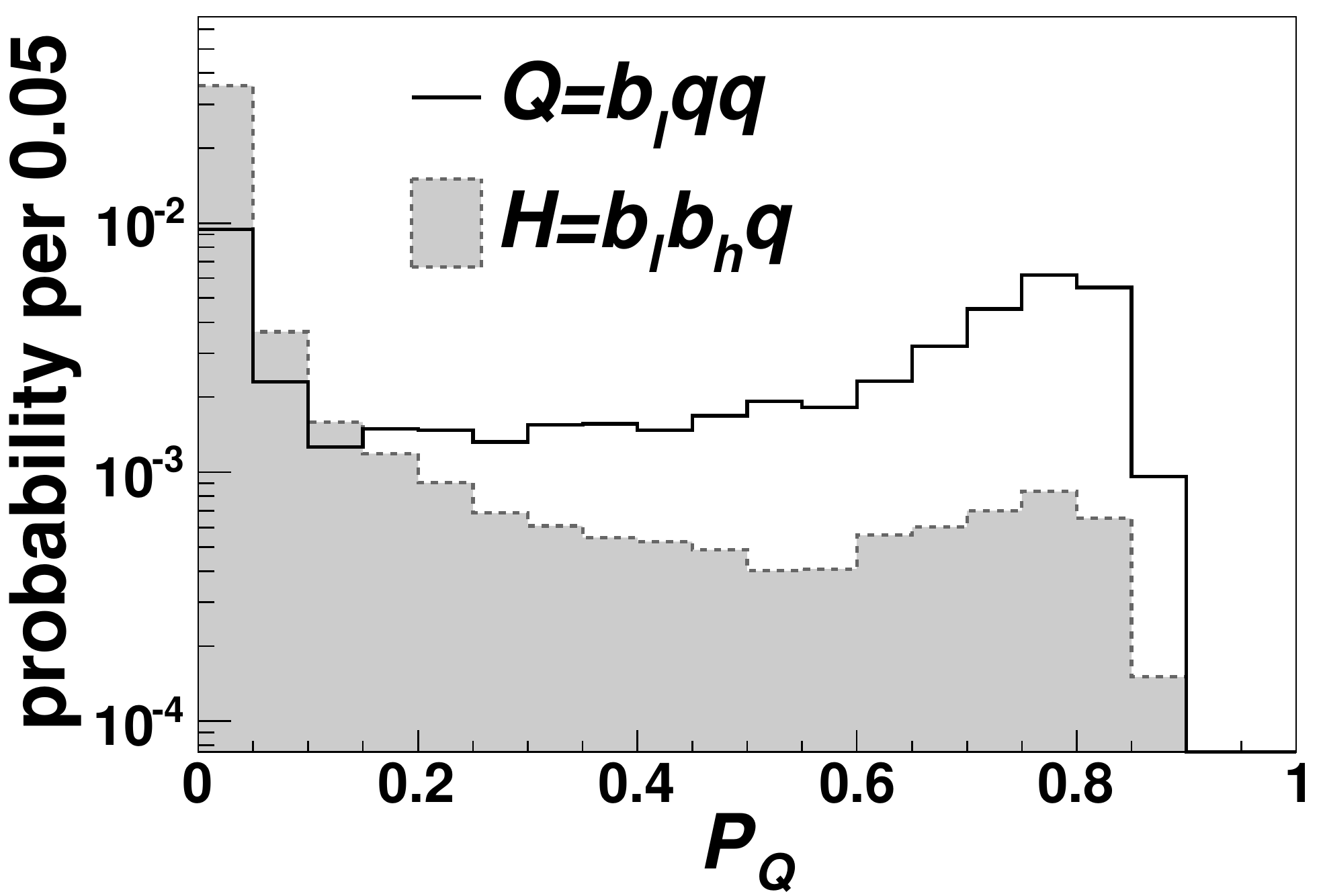}
\end{center}
\vspace{-0.6cm}
\caption{
Distribution in the a posteriori probability for a $Q$-type event, shown for $H$- and $Q$-type events.
}
\label{fig:fQ}
\end{figure}

The inclusion of the rarer $Q$ events in the likelihood can distort the reconstruction of
the more common case, the $H$ events. But this risk is mitigated when the likelihood
contains enough information to distinguish between the two cases on an event-by-event basis. 
To demonstrate that, we calculate the a posteriori probability that a matchable event 
is of type $Q$ as:
\begin{equation}
P_Q=\frac{f_Q P\left(d\mid a,Q\right)}{(1-f_Q) P\left(d\mid a,H\right)+f_Q P\left(d\mid a,Q\right)}
\end{equation}
As Fig.~\ref{fig:fQ} demonstrates the separation between the two cases is quite good.
This separation is mostly due to the $b$-tagging discriminants.
It is also useful to check the modeling of $P_Q$ against collider data, 
as all the terms in $P\left(d\mid a\right)$ also appear in $P_Q$.

%
\subsubsection{Scaling the proxy}
\label{sec:alpha}
\begin{figure}[htbp]
\centering
\begin{minipage}{.48\textwidth}
  \centering
  \includegraphics[width=\linewidth]{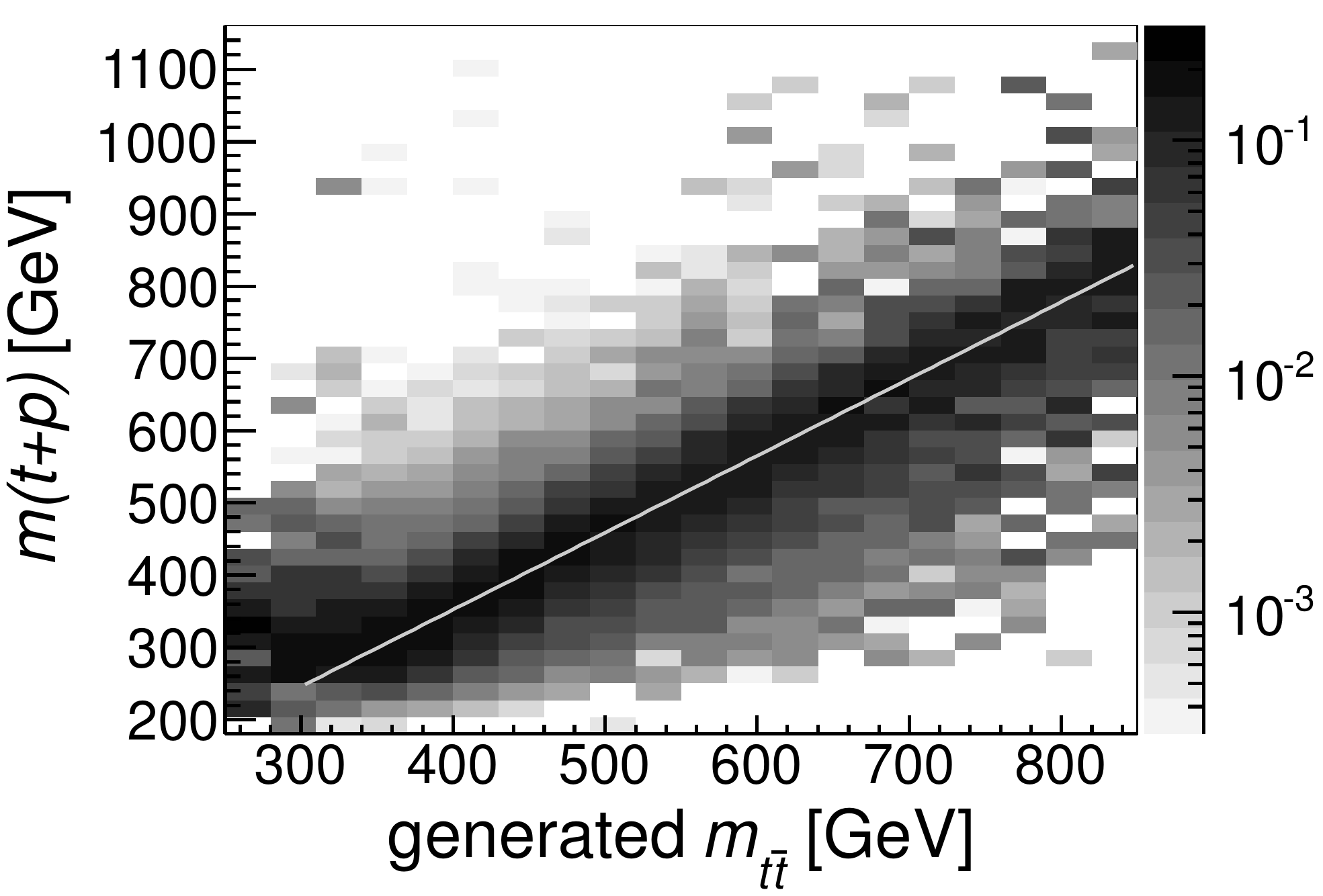}
  \caption{The distribution of the generated and reconstructed \ttbar\ invariant masses, without any scaling of the proxy, 
    for all selected events. The light-gray line shows a fit to the peak position of the reconstructed mass, \mpeak. }
  \label{fig:target}
\end{minipage}%
\hspace{.04\textwidth}%
\begin{minipage}{.48\textwidth}
  \centering
  \includegraphics[width=\linewidth]{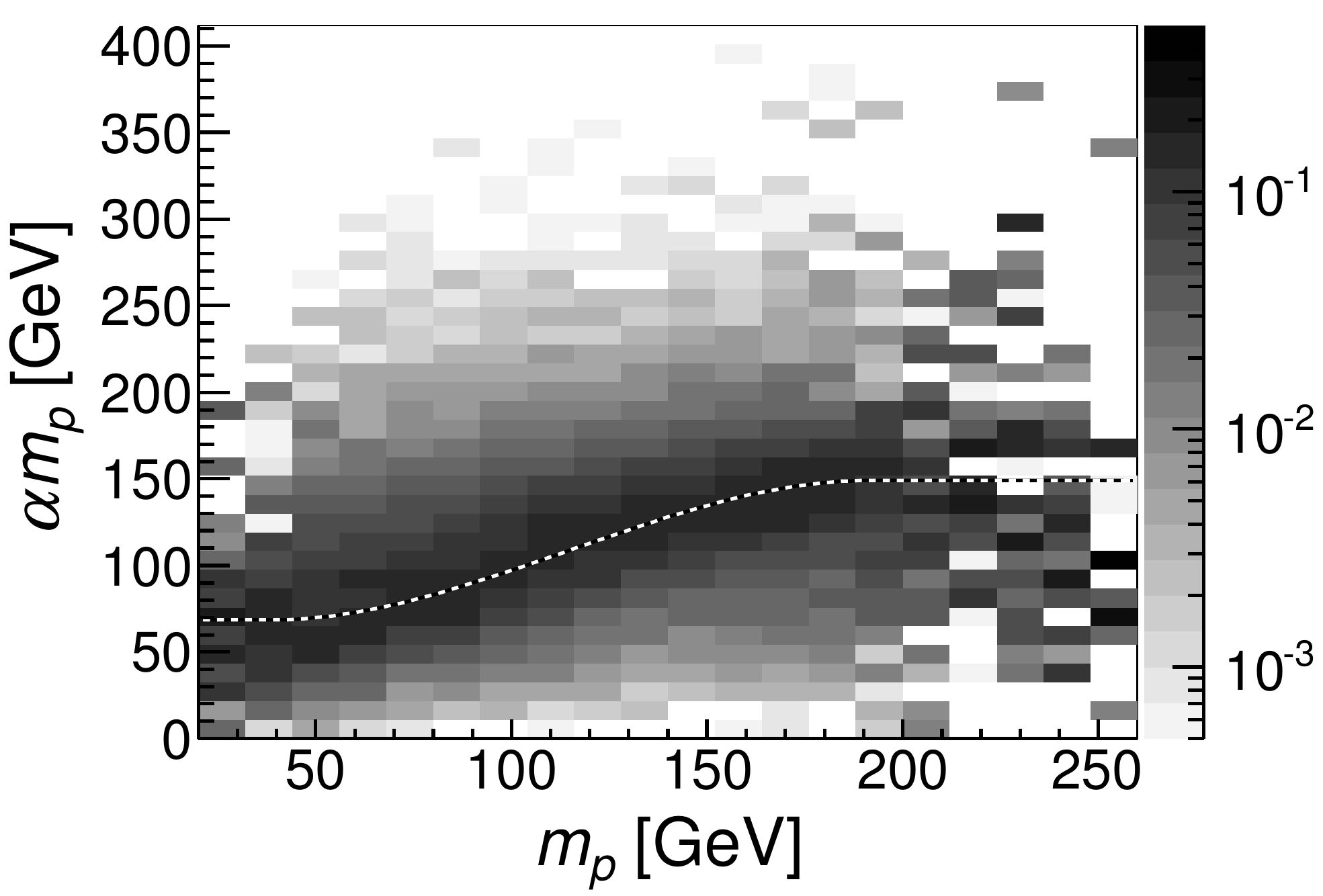}
  \captionof{figure}{The distribution of the proxy mass before and after scaling by $\alpha$, for all selected events.
    The dashed black and white curve shows a fit to the peak position of $\alpha$, $\ahat\left(\mproxy\right)$.}
  \label{fig:alpha_fit}
\end{minipage}
\end{figure}

Given a specific jet to quark assignment we have a candidate for the leptonic top $t$ with the energy $E_t$, momentum
$\vec{P_t}$ and invariant mass $m_t=\sqrt{E_t^2-\vec{P_t}^2}$ and a proxy $p$ for the hadronic top with the energy 
$E_p$, momentum $\vec{P_p}$ and invariant mass $m_p=\sqrt{E_p^2-\vec{P_p}^2}$.  
Since the proxy tends to underestimate the 4-vector of the hadronic top quark, 
the invariant mass of these two objects, $m(t+p)$, is likely to underestimate the generated invariant mass 
of the \ttbar\ system, \gmtt, as shown in  Fig.~\ref{fig:target}.
Additional scaling can be applied to the proxy 4-vector to partially correct for this underestimation.
Furthermore, since the reconstructed proxy mass, \mproxy, indicates the size of the underestimation in each event,
this scaling can  be parametrized as a function of \mproxy. 
 
For each simulated event, we define the ideal scaling of the proxy 4-vector, $\alpha$, as the
scale that will bring the reconstructed $m(t+p)$ to the peak position\footnote{It is tempting to define the
ideal as $m(t+p)=\gmtt$, which will also calibrate the reconstructed \mttbar. But it is more
important to reduce the scatter in the reconstructed \mttbar, and needlessly introducing the calibration
lowers the effectiveness of the derived scaling.} of the reconstructed mass, \mpeak (see Fig.~\ref{fig:target}).
Since \mpeak\ is a function of \gmtt, this scale is unavailable in collider data. 
Instead, we reconstruct events using a scale \ahat\ which is an estimate of $\alpha$ based on 
the observable \mproxy.

To derive this estimate, we solve for $\alpha$ in simulated events, 
which results in a quadratic equation:
\begin{equation}
\alpha^2m_p^2+2\alpha\left(E_tE_p-\vec{P_t}\vec{P_p}\right)+\left(m_t^2-\mpeak^2\right)=0.
\end{equation}
We then plot, in Fig.~\ref{fig:alpha_fit}, the two-dimensional distribution of the proxy mass scaled by $\alpha(\gmtt)$ 
and the unscaled \mproxy.
From this distribution we parametrize the most probable value of $\alpha$  as a function of \mproxy\ to find
our estimated \ahat.
The parametrization of $\ahat\left(\mproxy\right)$ was chosen from polynomial functions that were constrained so that
the scaled mass, $\ahat\mproxy$, is non-decreasing~\footnote{This is enforced only at the
edge of the distributions. Though the middle of the function was allowed to decrease,
the best-fit function does not do so.}.
Finally, we construct the invariant mass of the \ttbar\ system from the sum of 
the 4-vector of the proxy, scaled by $\ahat\left(\mproxy\right)$, and the 4-vector of the leptonic top candidate.

%
\subsubsection{Averaging the assignments}
\label{sec:averaging}
The most significant improvement is from considering more than one jet assignment.
The algorithms described so far considered only the most likely assignment,
the one that minimizes the \chisq\ in Eq.~\ref{eq:chisq} or that 
maximizes $P\left(a\mid d\right)$ in Eq.~\ref{eq:prob}.
But we can also use all the possible assignments weighted by their a posteriori probabilities.
For example:
\begin{equation}
\amtt=\sum\limits_a m_{t\bar{t}}^a P\left(a\mid d\right).
\end{equation}
These averaged reconstructions tend to have the advantage of a spread lower than that of
the single-assignment reconstructions, and the disadvantage of a lower response.
Here we define the ``response'' for an observable as the derivative of the average reconstructed value as a function of
the true, generated value and the ``spread'' as the RMS of the distribution of the reconstructed value 
for a fixed true, generated value. 

%
%
\section{\boldmath Performance}
\label{sec:performance}
\subsection{Definition of the figure of merit}
\label{sec:FOM}
To compare the performance of different reconstruction algorithms, 
we require an appropriate figure of merit.
Algorithm performance is usually quantified by summarizing the distribution of the difference 
(or the ratio) between the reconstructed and generated observable into its RMS, 
or into the width of a Gaussian fit to the core of the distribution.
However, this quantification
presumes that the reconstruction is unbiased and centered around the true value.
For the reconstruction algorithms discussed 
here\footnote{And also for other \ttbar\ reconstruction algorithms for \lpgefj\ events.} 
the difference distributions are intrinsically bimodal, 
since the performance differs for matchable and unmatchable events. 

For matchable events, the reconstruction typically has a response that is close to one 
and a narrow spread, while for the unmatchable events it typically has a low response and a wide spread.
Hence the average reconstruction is biased, while the peak position is almost unbiased,
and the reconstruction can not be calibrated so it is both unbiased and peaks at the generated value.

\begin{figure}[htbp]
\begin{center}
\includegraphics[width=1\linewidth]{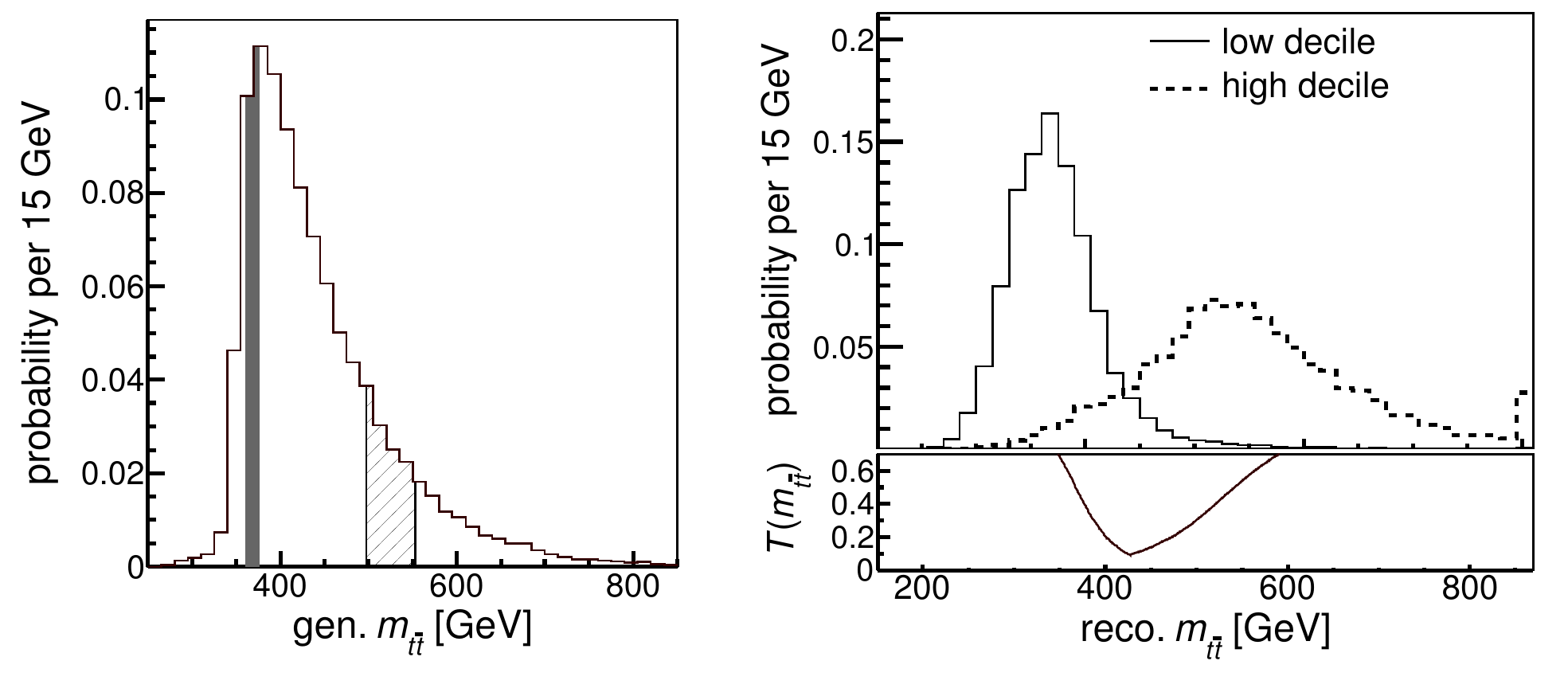}
\end{center}
\begin{picture}(0,0)(0,0)
\put (150,191){\subfloat[][]{\label{subfig:fom_a}}}
\put (360,191){\subfloat[][]{\label{subfig:fom_b}}}
\end{picture}
\vspace{-1.2cm}
\caption{
An example of the minimal overlap as a figure of merit for a reconstruction.
The events of $f_L$ are shaded in (a) and shown with the solid curve in (b);
the events of $f_H$ are hatched in (a) and shown with the dashed curve in (b).
Overflows are shown in the edge bins.
The offset is $s=0.1$, the point of minimal overlap is at $\mttbar=427\GeV$,
as shown in the lower panel of (b), and $\Nsigma=2.65$.
}
\label{fig:FOM}
\end{figure}
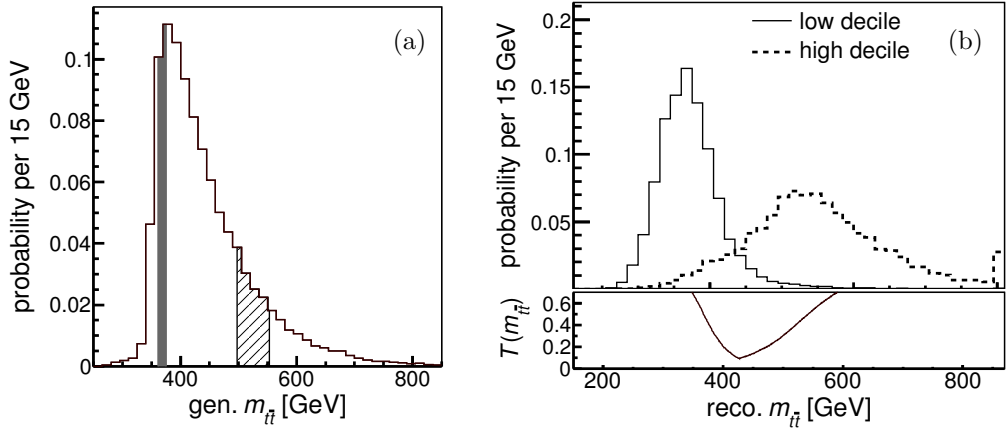

To quantify the quality of the reconstruction without relying on the properties of
its calibration, we contrast the reconstructed observable for two categories of events, defined by the 
quantiles of the generated observable. This is demonstrated in Fig.~\ref{fig:FOM}.
Each category contains 10\% of the events, and they are defined according to an offset, 
$s$, so that one category is generated between the $s$ and $s+0.1$ quantiles and the other
between the $0.9-s$ and $1-s$ quantiles (see Fig.~\ref{subfig:fom_a} where the 2nd and 9th
deciles are used). 
The FOM quantifies how well the reconstruction separates these two categories.

We denote the distributions of the reconstructed observable for these categories $f_L$ and $f_H$.
An example is shown in Fig.~\ref{subfig:fom_b}.
Were these distributions Gaussian and identical, it would be natural to quantify the separation in terms
of \Nsigma, the number of standard deviations between their peaks.
To generalize this concept to arbitrary distributions and to focus on the possible misclassification
of events between the two categories, we define $T(x)$ as the overlap between 
these distributions at observable value $x$ and the minimal overlap $M$:
\begin{align}
M = \min_{x} T(x), && T(x) = \max\left( \int_x^{+\infty} f_L(x')\dd x', \int_{-\infty}^x f_H(x')\dd x' \right).
\end{align}
These too are shown in Fig.~\ref{subfig:fom_b}.
Smaller $M$ values indicate less misclassification and hence better performance of the reconstruction
algorithm.

\begin{figure}[htbp]
\centering
\begin{center}
  \includegraphics[width=0.46\linewidth]{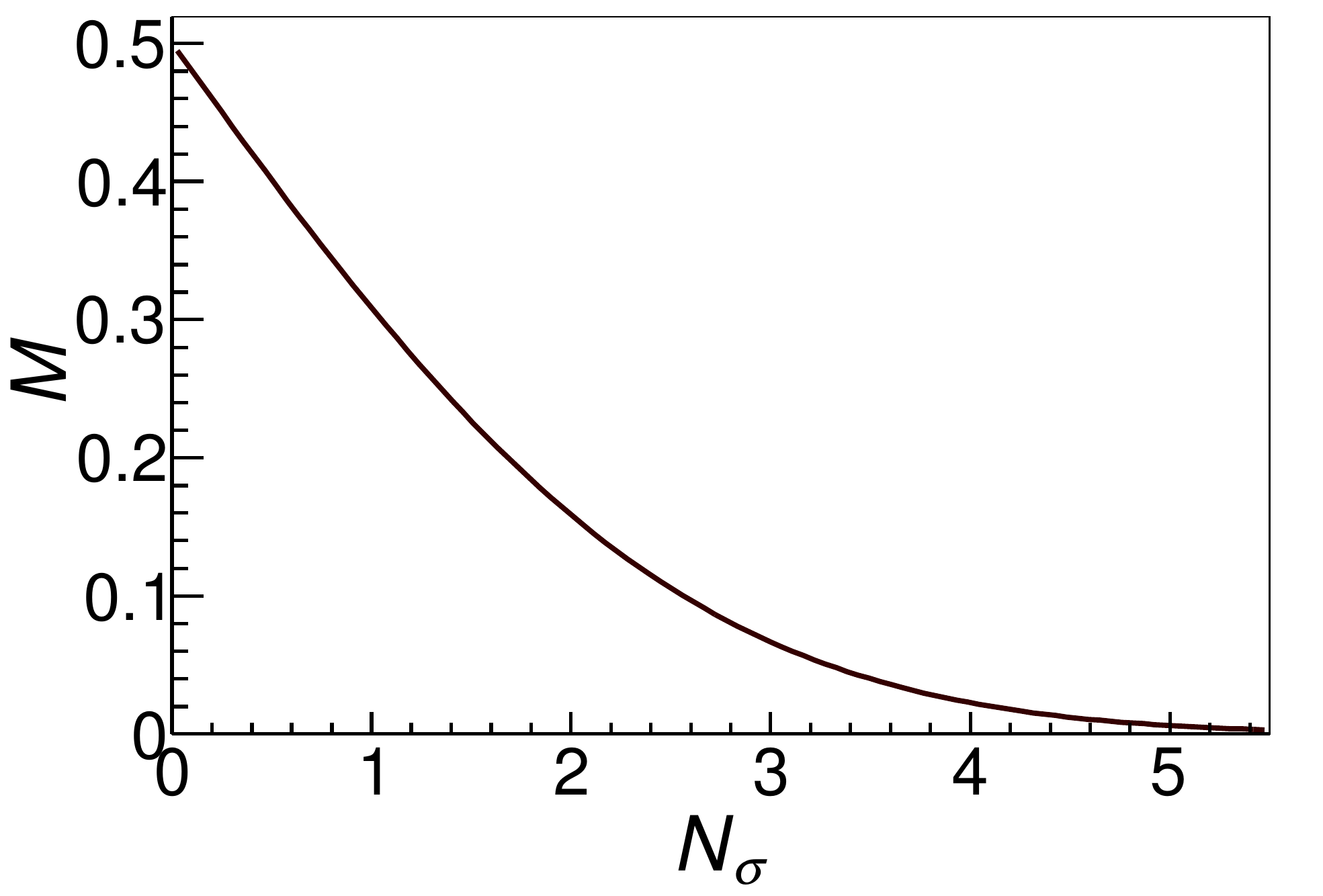}
  \hspace{0.02\linewidth}
  \includegraphics[width=0.46\linewidth]{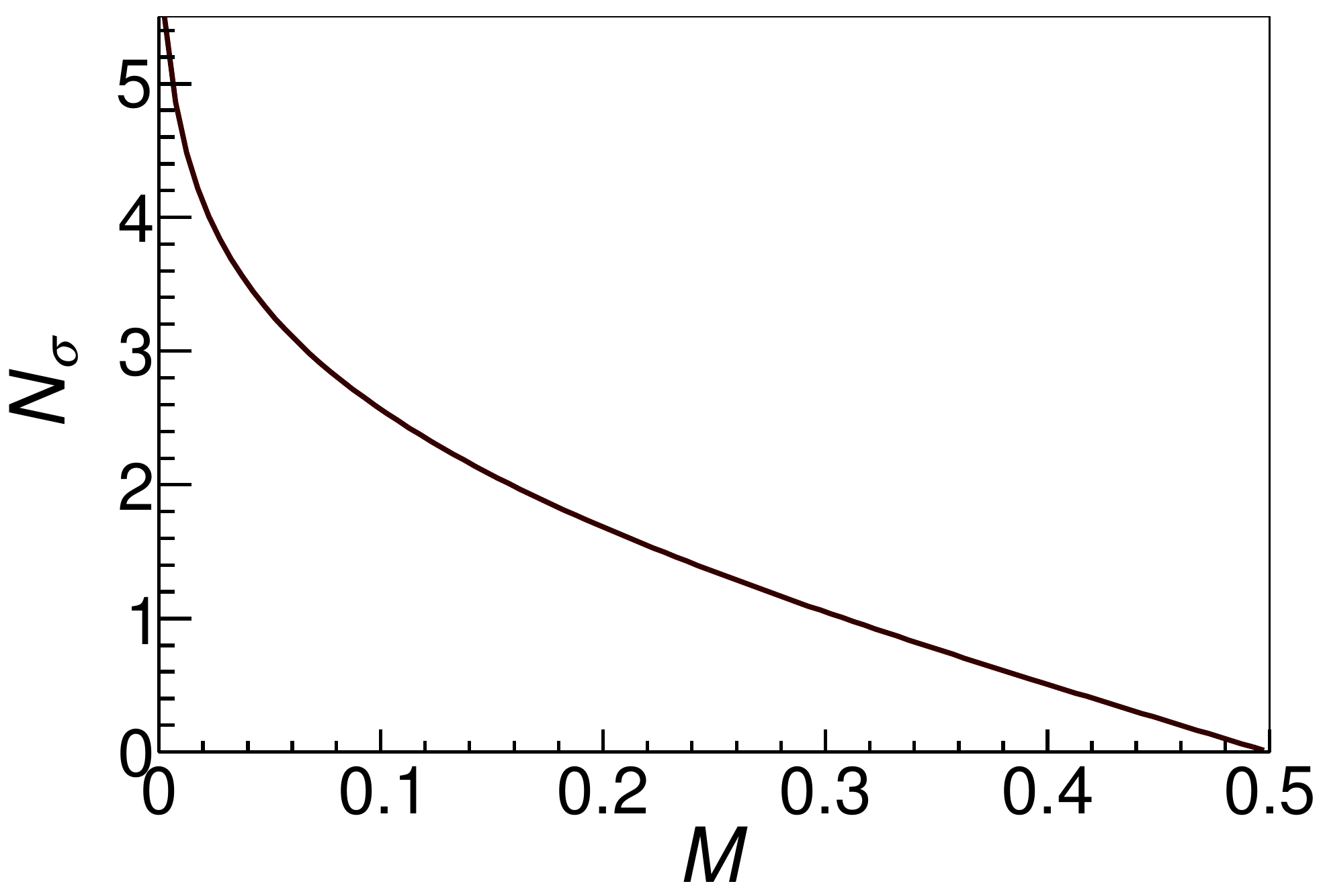}
\end{center}
\begin{picture}(0,0)(0,0)
\put (-37,143){\subfloat[][]{\label{subfig:sas_a}}}
\put (160,143){\subfloat[][]{\label{subfig:sas_b}}}
\end{picture}
\vspace{-1.2cm}
\caption{The minimal overlap and the number of $\sigma$s.}
\label{subfig:sep_and_sigma}
\end{figure}
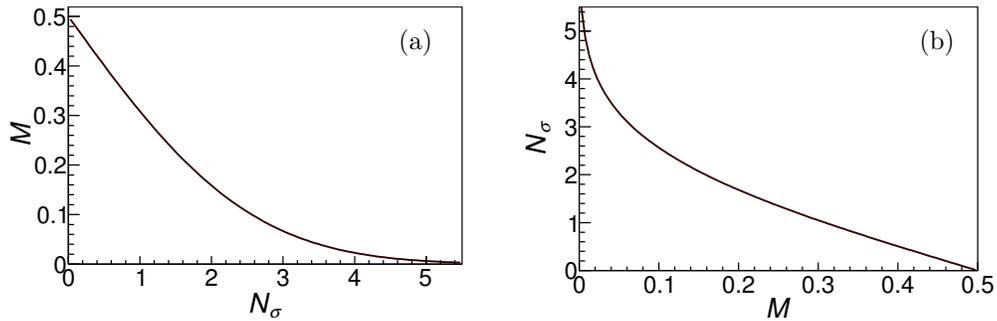

We can translate $M$ to the more familiar ``number of $\sigma$s'' by considering $M$ for two Gaussian
distributions of width one, whose means are separated by \Nsigma:
\begin{equation}
M\left(\Nsigma\right) = \int_{\frac{1}{2}\Nsigma}^{\infty} G(x)\dd x = \frac{1}{2}\left(1-\erf\left(\frac{\Nsigma}{2\sqrt{2}}\right)\right),
\end{equation}
where $G$ is the normal distribution (see Fig.~\ref{subfig:sas_a}).
By inverting this relationship (see Fig.~\ref{subfig:sas_b}), 
we can present the minimal overlap in terms of \Nsigma.

This FOM has another, incidental advantage.
Unlike RMS values, it can be interpreted without referring to the width and shape
of the expected generated distribution.
%
\subsection{Comparison of the algorithms}
\label{sec:comparison}

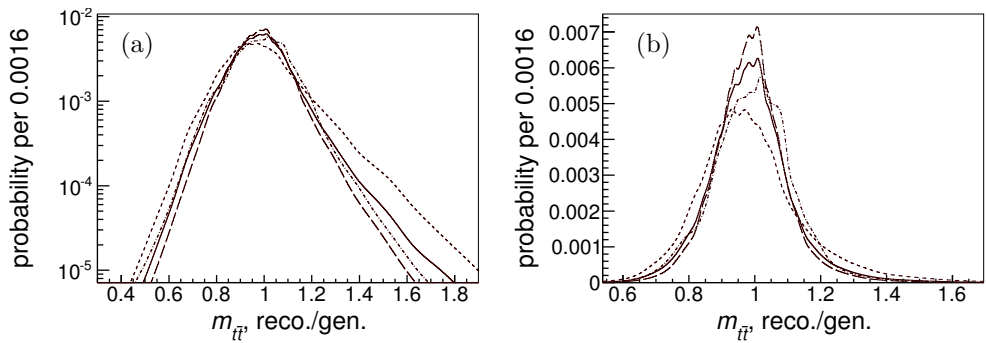
\begin{figure}[htbp]
\begin{center}
\includegraphics[width=0.48\linewidth]{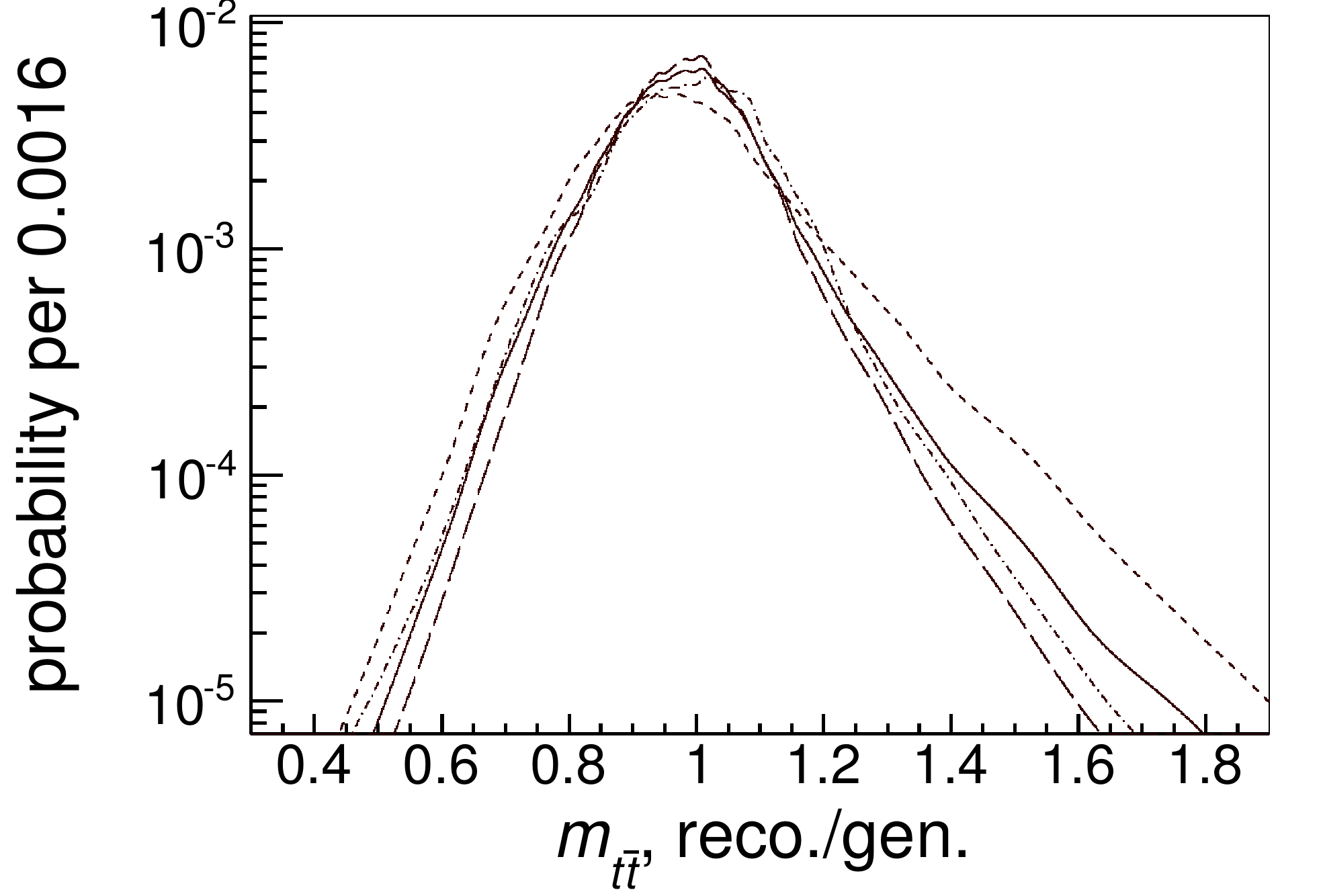}
\includegraphics[width=0.48\linewidth]{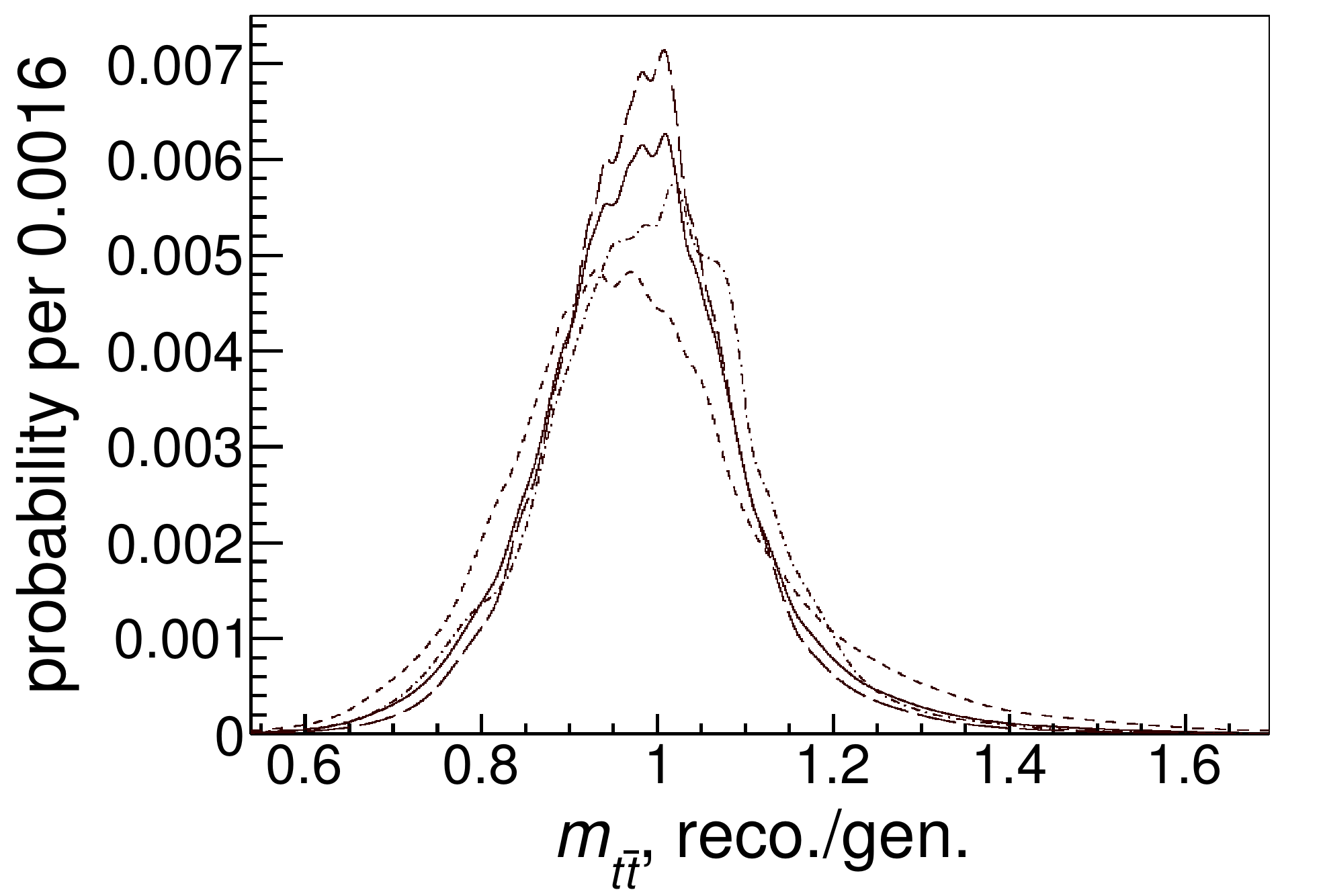}
\end{center}
\begin{picture}(0,0)(0,0)
\put ( 50,150){\subfloat[][]{\label{subfig:mtt_res_a}}}
\put (245,150){\subfloat[][]{\label{subfig:mtt_res_b}}}
\end{picture}
\vspace{-1.2cm}
\caption{
Resolution in \mttbar.
The y axis in the left-hand plots is on a logarithmic scale, while the right-hand plots show the peak region
on a linear scale.
Events where one of the jets from \Whboson\ decay is lost (case $H$) are shown by the long-dashed curves,
events where the hadronic $b$ jet is lost (case $Q$) are shown by the dashed-dotted curves,
unmatchable events are shown by the dashed curves, 
and the solid curves show all events.
}
\label{fig:mtt_res}
\end{figure}

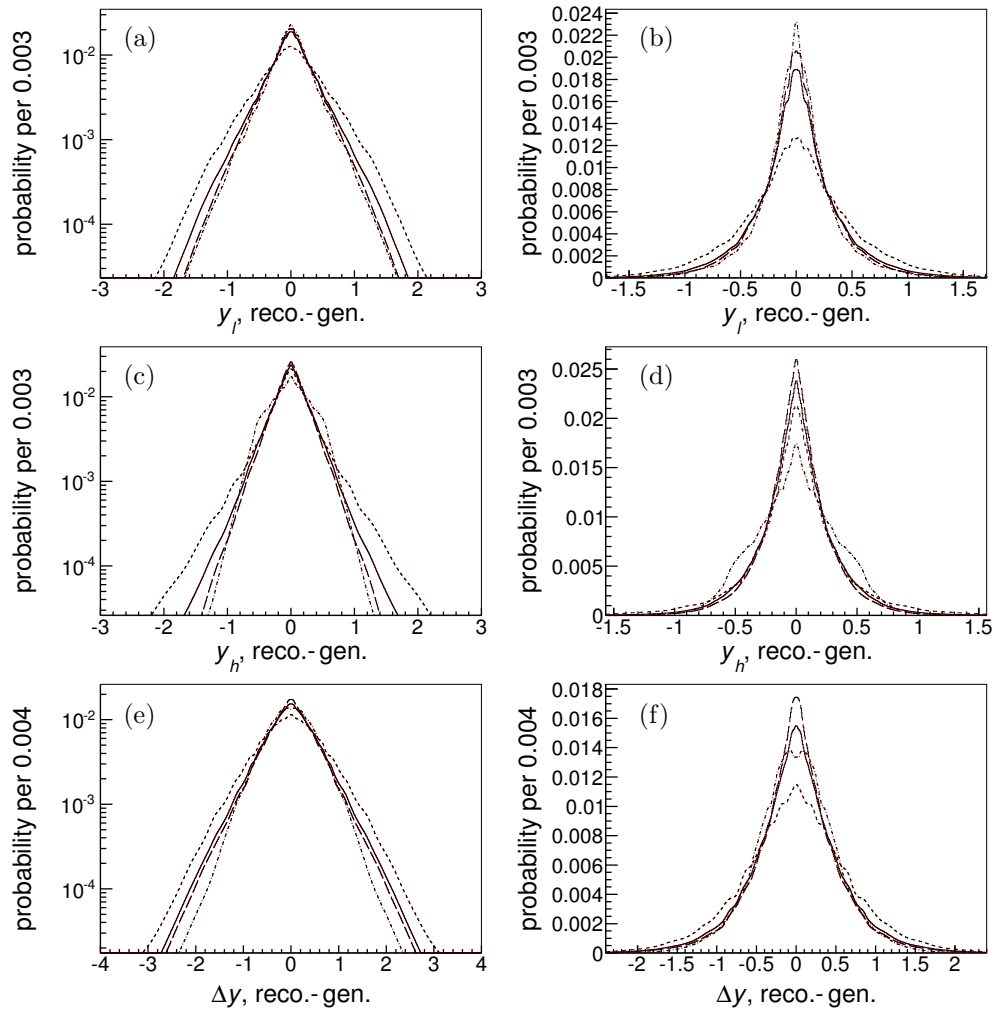
\begin{figure}[htbp]
\begin{center}
\includegraphics[width=0.48\linewidth]{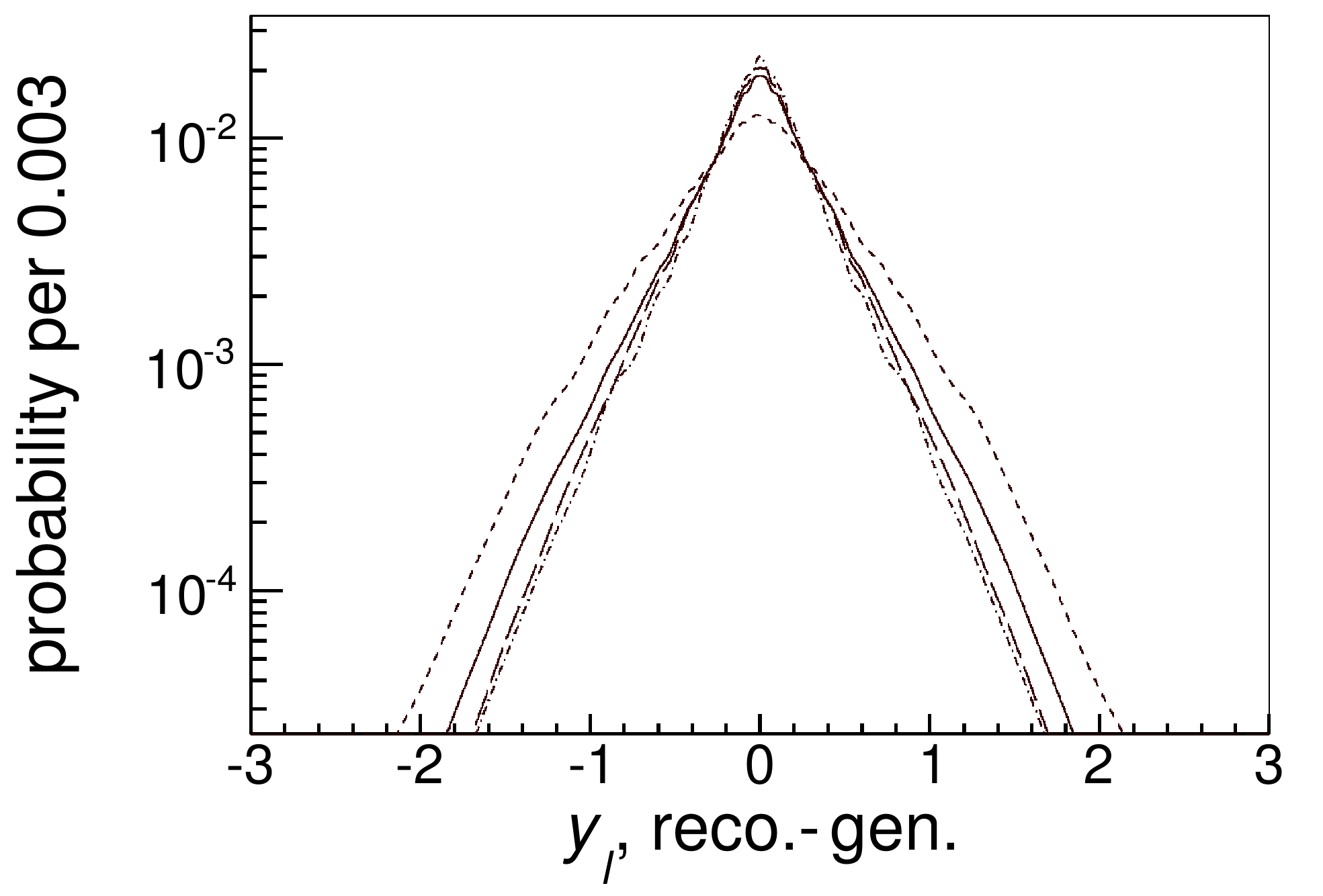}
\includegraphics[width=0.48\linewidth]{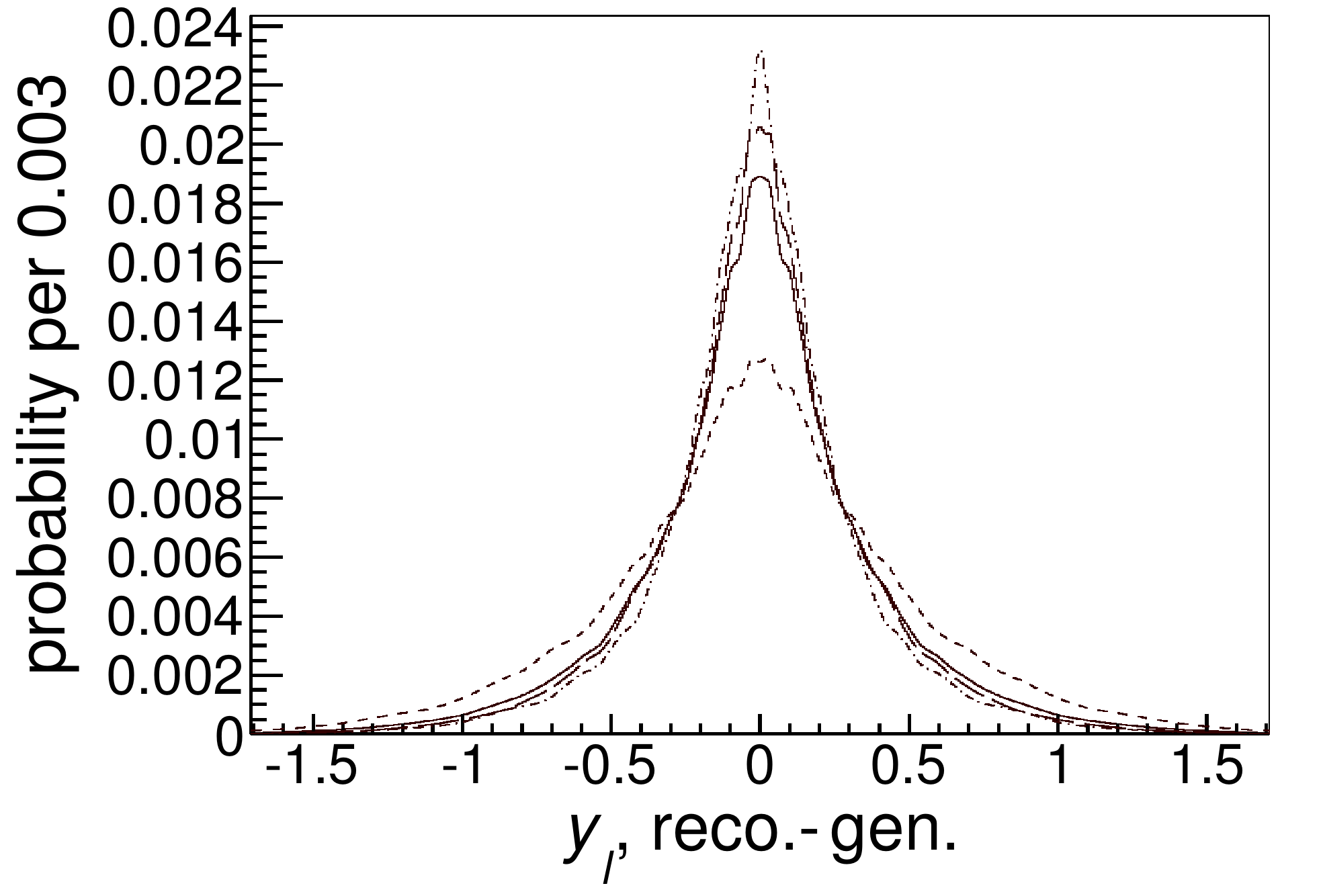}\\
\includegraphics[width=0.48\linewidth]{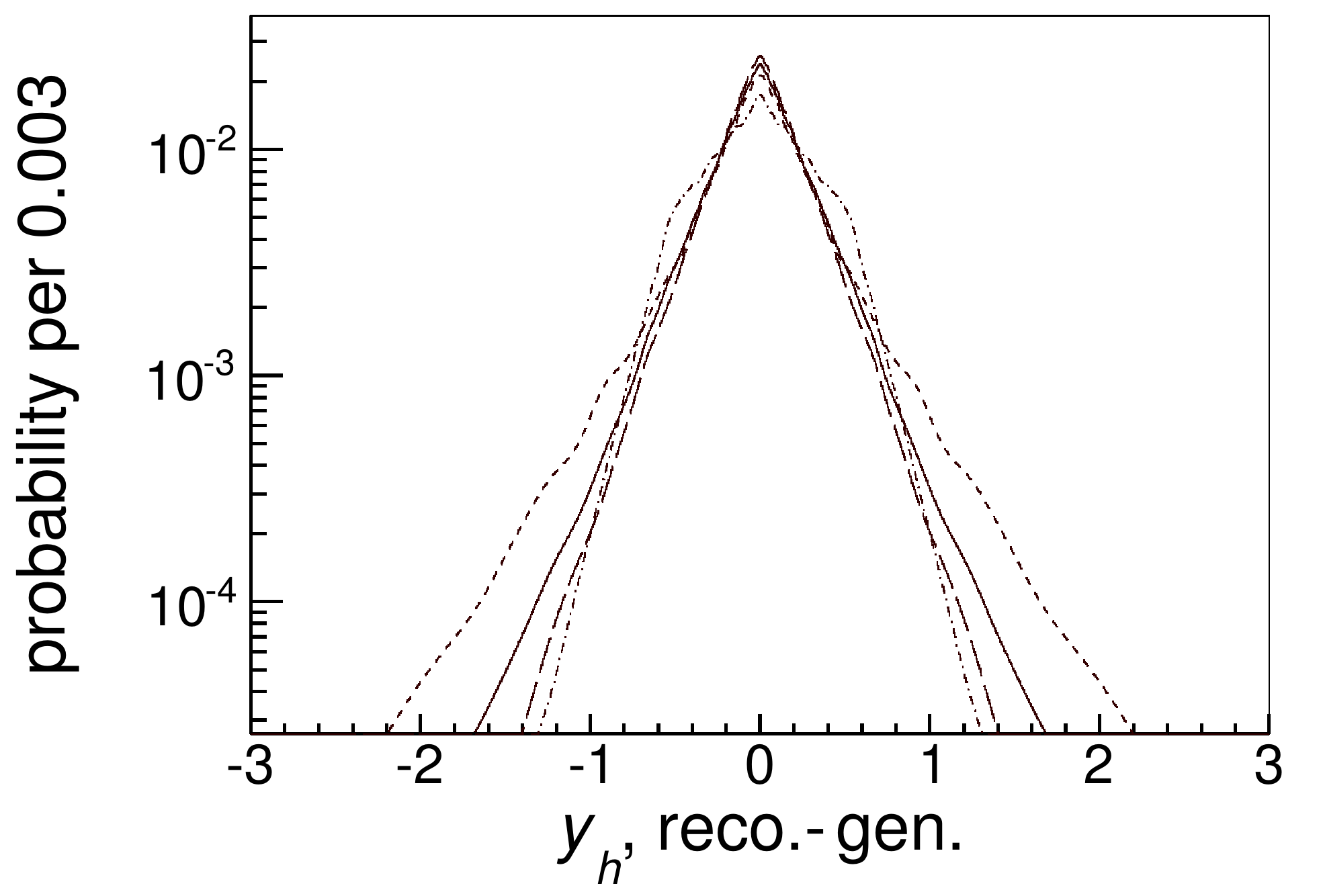}
\includegraphics[width=0.48\linewidth]{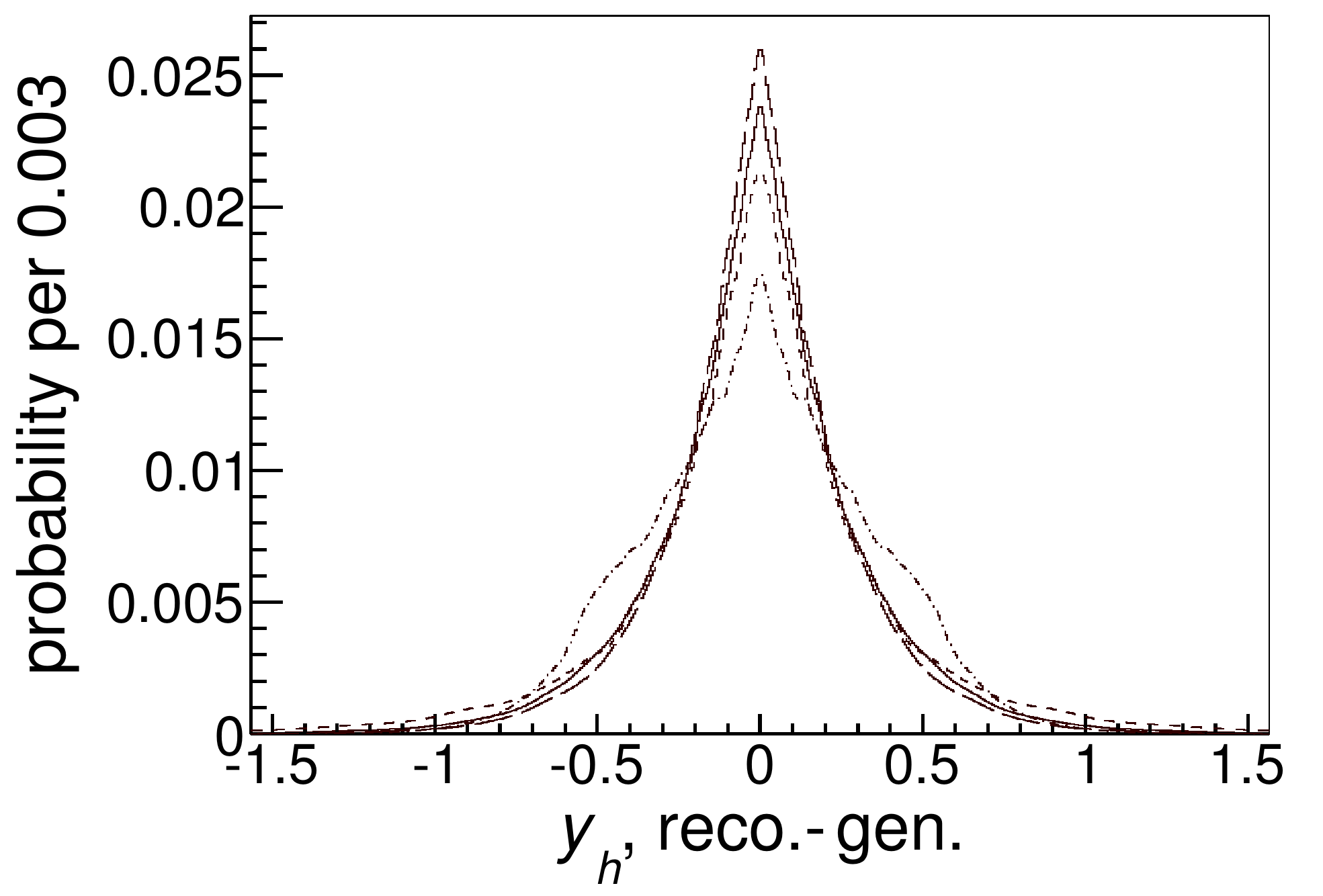}\\
\includegraphics[width=0.48\linewidth]{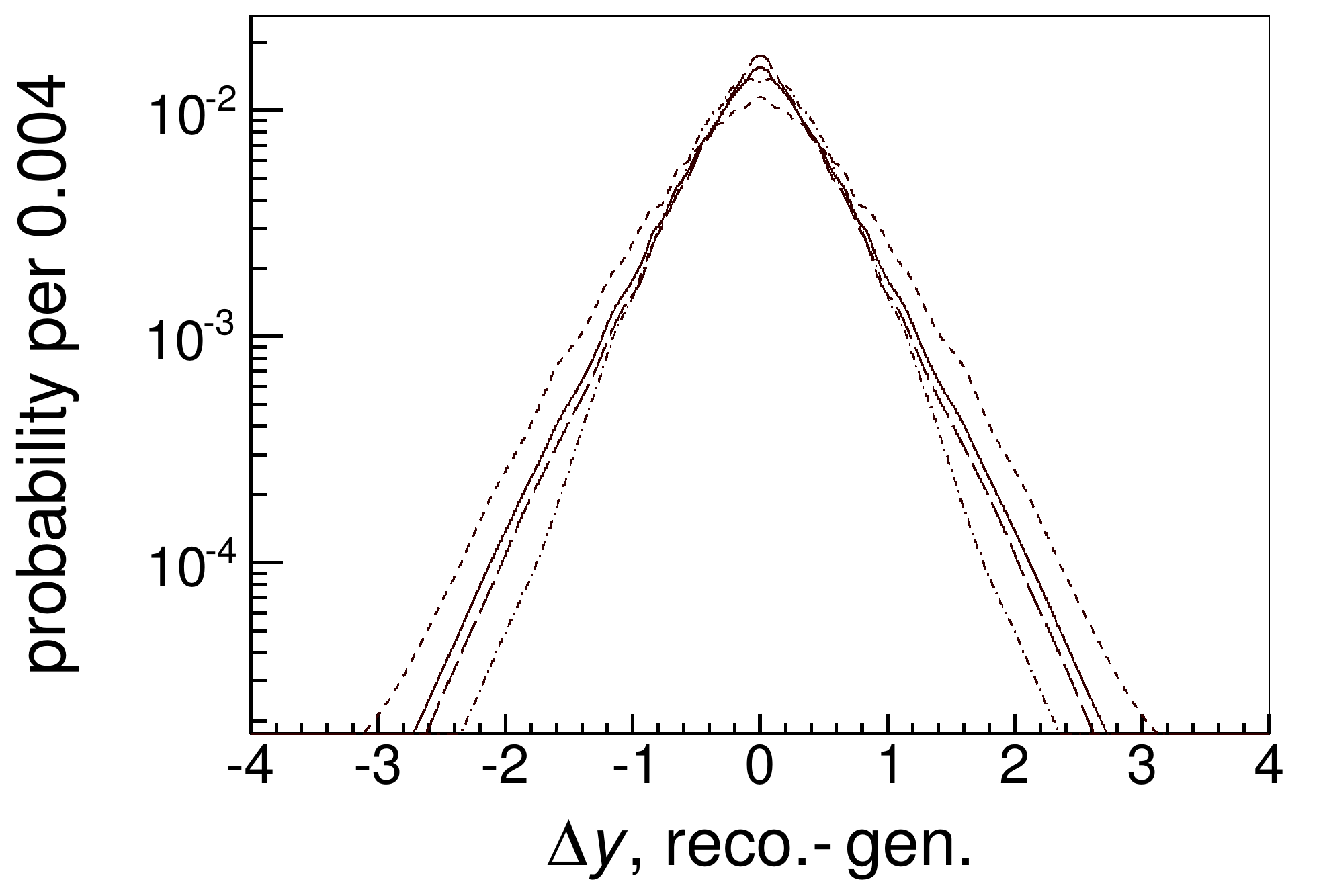}
\includegraphics[width=0.48\linewidth]{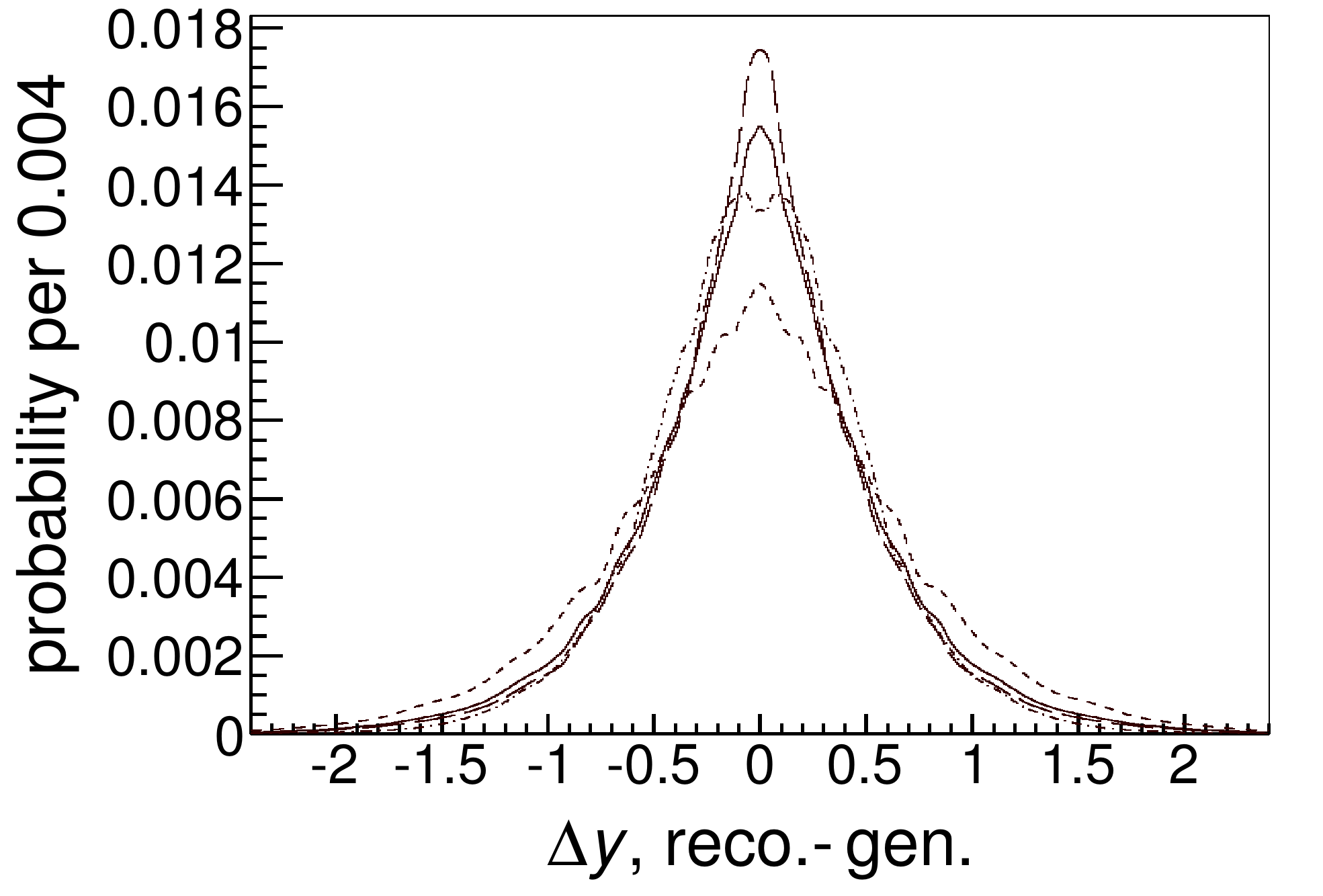}
\end{center}
\begin{picture}(0,0)(0,0)
\put ( 50,406){\subfloat[][]{\label{subfig:yres_a}}}
\put (245,406){\subfloat[][]{\label{subfig:yres_b}}}
\put ( 50,278){\subfloat[][]{\label{subfig:yres_c}}}
\put (245,278){\subfloat[][]{\label{subfig:yres_d}}}
\put ( 50,150){\subfloat[][]{\label{subfig:yres_e}}}
\put (245,150){\subfloat[][]{\label{subfig:yres_f}}}
\end{picture}
\vspace{-1.2cm}
\caption{
Resolution in top-quark rapidity on the leptonic side (a,b) and the proxy side (c,d), and in \dy\ (e,f).
The y axis in the left-hand plots is on a logarithmic scale, while the right-hand plots show the peak region
on a linear scale.
Events where one of the jets from \Whboson\ decay is lost (case $H$) are shown by the long-dashed curves,
events where the hadronic $b$ jet is lost (case $Q$) are shown by the dashed-dotted curves,
unmatchable events are shown by the dashed curves, 
and the solid curves show all events.
As we expect symmetric resolution functions, we construct all curves to be symmetric.
}
\label{fig:yres}
\end{figure}

Figures~\ref{fig:mtt_res} and~\ref{fig:yres} compare the reconstruction of different classes of 
events with the new algorithm. For ease of display, a rough linear calibration of \amtt\ is used when displaying
the resolutions of the partial reconstruction algorithm.
Both classes of matchable events (case $H$ and case $Q$) are reconstructed well, and the
reconstruction of unmatchable events is not much worse.
As $76(1)\%$ of the events are matchable, the reconstruction for all events 
is almost as good as for matchable events.
The reconstruction of the hadronic-top rapidity is especially weak for events of type $Q$,
indicating that a missing ``hadronic'' $b$ jet is more problematic than a missing
jet from \Whboson\ decay. 
The reconstruction of the leptonic-top rapidity is especially weak for unmatchable events,
since for most of these events the ``leptonic'' $b$ jet is lost.

\begin{figure}[htbp]
\begin{center}
\includegraphics[scale=.3]{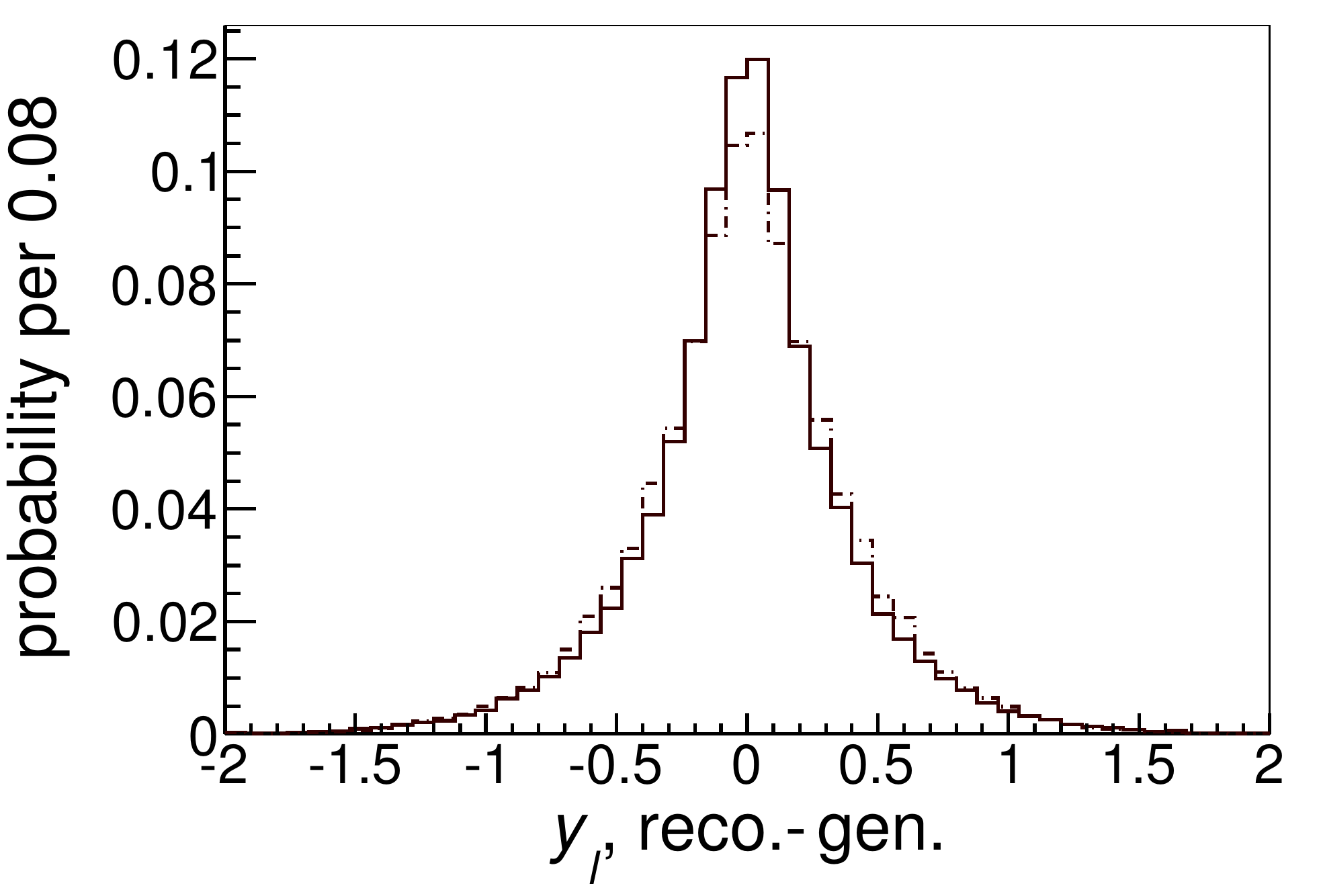}
\includegraphics[scale=.3]{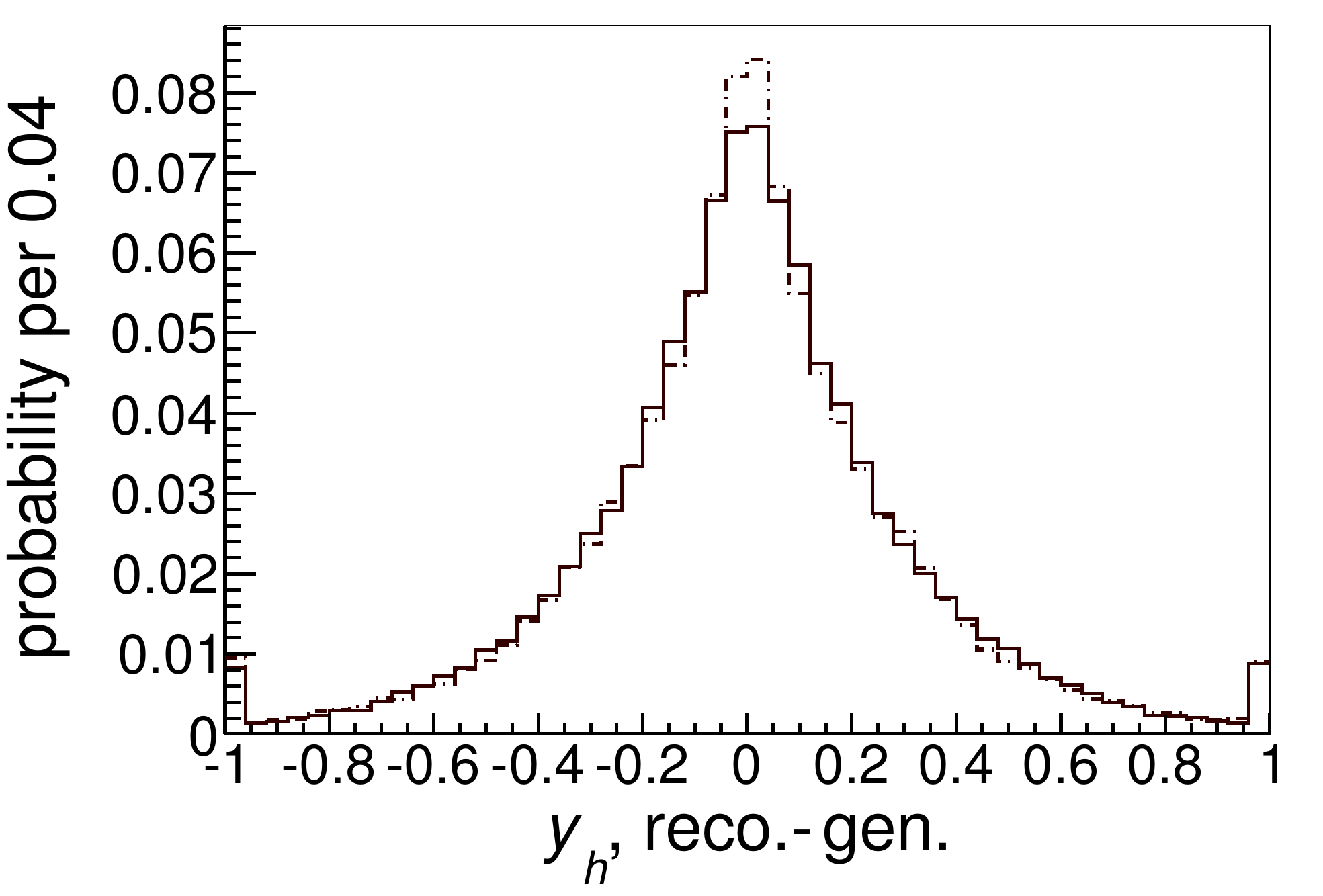}
\includegraphics[scale=.3]{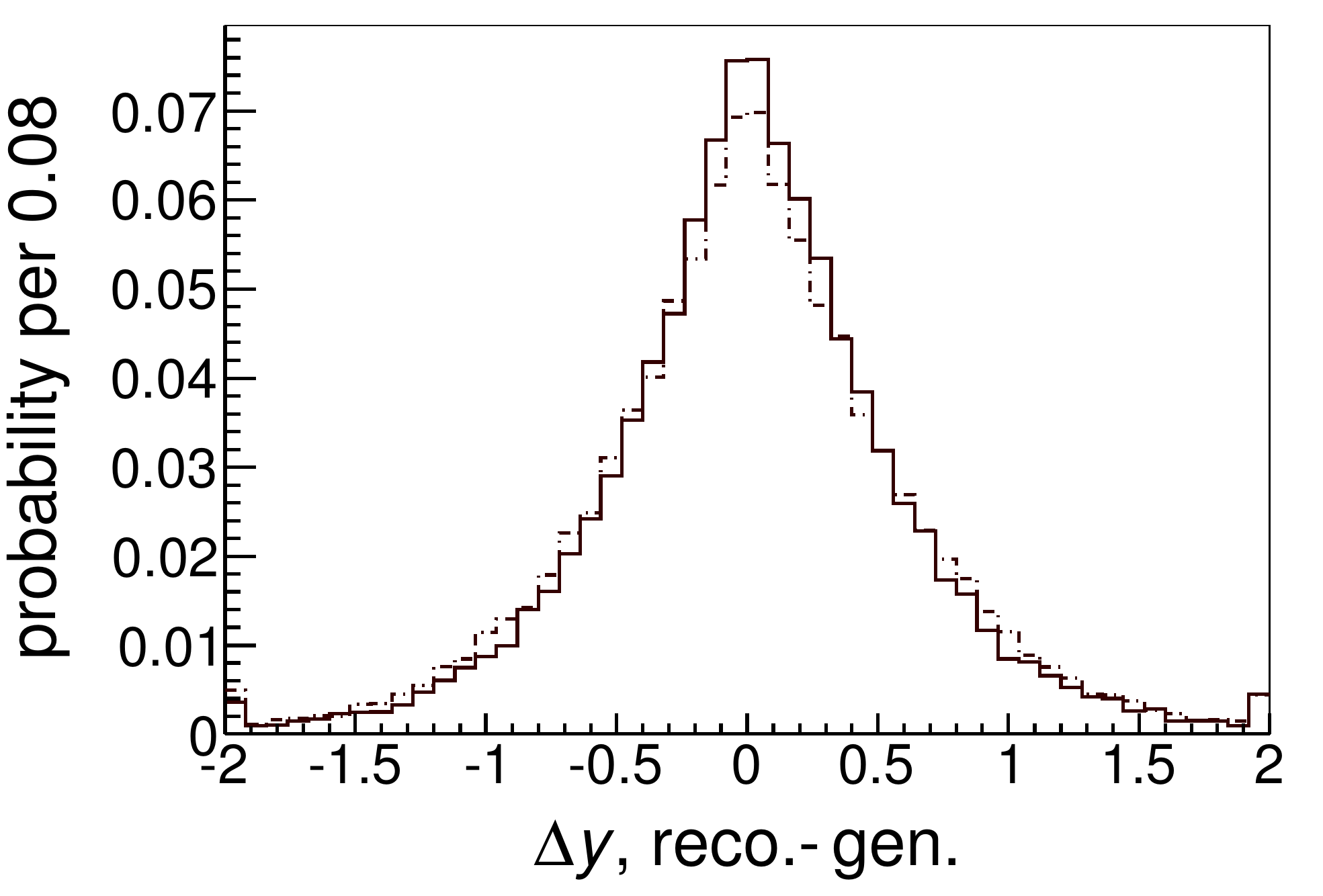}
\includegraphics[scale=.3]{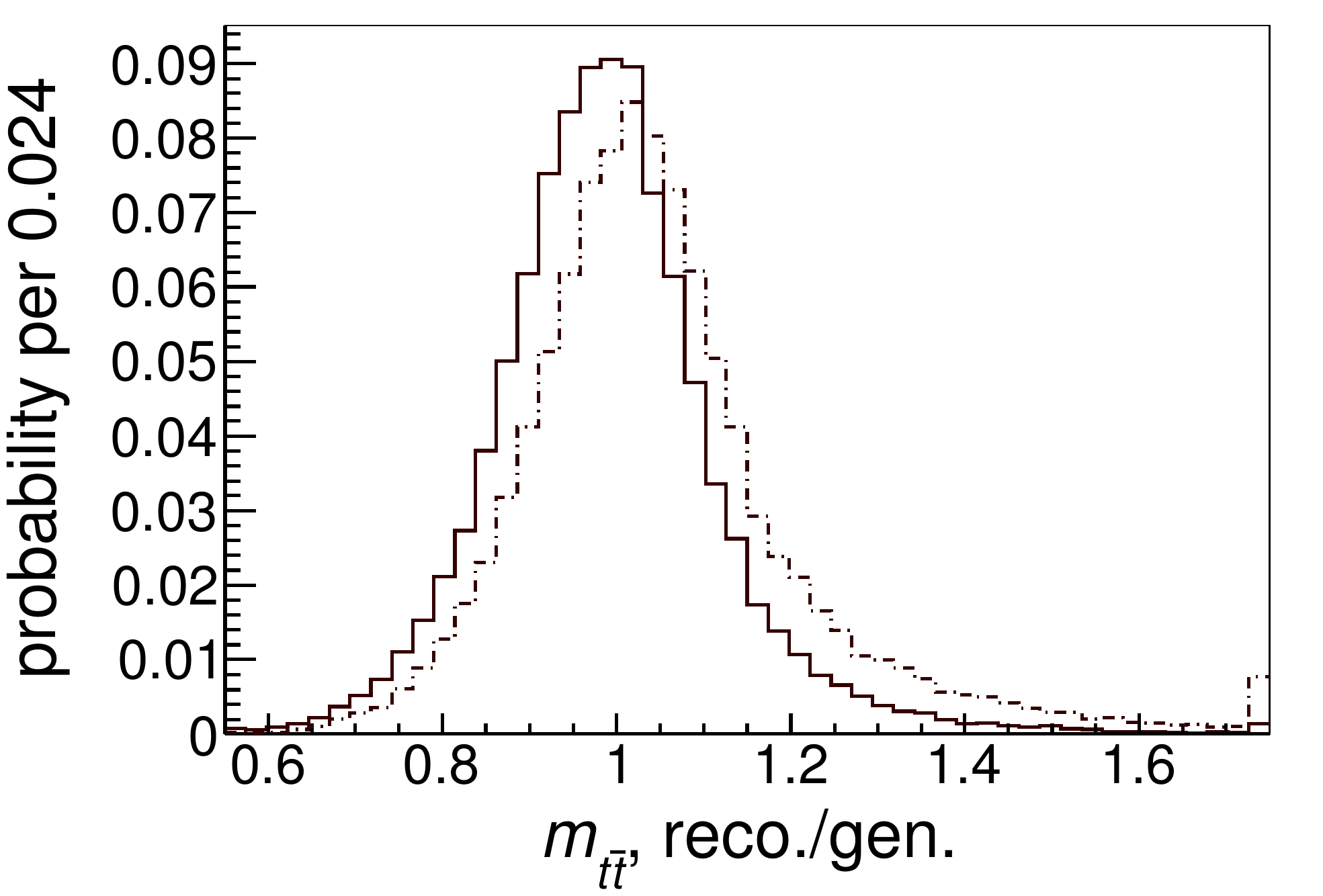}
\end{center}
\begin{picture}(0,0)(0,0)
\put (170,257){\subfloat[][]{\label{subfig:3vs4_a}}}
\put (345,257){\subfloat[][]{\label{subfig:3vs4_b}}}
\put (170,140){\subfloat[][]{\label{subfig:3vs4_c}}}
\put (345,140){\subfloat[][]{\label{subfig:3vs4_d}}}
\end{picture}
\vspace{-1.2cm}
\caption{
Resolution in (a) $y_l$, (b) $y_h$, (c) \dy, and in (d) \mttbar\ for
\lptj\ events (solid curve) and for \lpgefj\ events reconstructed 
with a kinematic fit algorithm~\cite{bib:hitfit} (dot-dashed curve).
In both cases, the weighted average of all assignments is used.
}
\label{fig:comp_3to4}
\end{figure}

No partial reconstruction algorithm was previously applied to $\ttbar\to\lpj$ events,
so we choose to compare the performance of the algorithm described in this \this\ to that of 
a kinematic fit algorithm that was used to fully reconstruct \lpgefj\ events~\cite{bib:hitfit}
in many top measurements (e.g. in Refs.~\cite{bib:hitfit_D0} and~\cite{bib:hitfit_CMS}).
As with the new algorithm, we can either use the most likely assignment from the kinematic fit algorithm
or use a weighted average of all assignments. The relative weight of each assignment
is $\exp\left(-\chisq/2\right)$, as in Ref.~\cite{bib:hitfit_D0}.

We compare the performance of the two algorithms for the ability to reconstruct the following observable:
the invariant mass of the \ttbar\ system (\mttbar), the rapidity of the leptonically decaying top quark ($y_l$), 
the rapidity of the hadronically  decaying top quark ($y_h$) and the rapidity difference (\dy$=y_l-y_h$). 
The distributions of the differences and ratio between reconstructed and generated 
observables for these two algorithms, shown in Fig.~\ref{fig:comp_3to4}, illustrate 
that the partial reconstruction provides a performance similar in quality to that of the full reconstruction. 

Table~\ref{tab:perf} uses the FOM introduced in Section~\ref{sec:FOM} to quantitatively compare 
the performance of the two algorithms. 
As the generated distributions differ between the \lptj\ and the \lpgefj\ samples,
there is some arbitrariness in such a comparison. 
To quantify this arbitrariness, for the \lpgefj\ samples each FOM was evaluated twice, once 
using the quantiles found in the \lpgefj\ sample and once using the quantiles found in the \lptj\ sample.
 
Table~\ref{tab:perf} also lists the performance of simpler versions of the new algorithm, 
corresponding to Sections~\ref{sec:chi2j},~\ref{sec:lhoodj},~\ref{sec:alpha}, and~\ref{sec:averaging}.
A constant offset, $s$, was chosen for each observable (\mttbar, $y_l$, $y_h$ and \dy). 
The offsets were chosen so the resulting \Nsigma\
values are $\approx 2$, a level of separation 
where further improvements are still useful (see Fig.~\ref{subfig:fom_b}).
Though the tail behavior of the reconstructions varies,
the variations are limited to a fraction of events much smaller than
the 10\% we consider in each category. Thus the choice of offsets 
has little effect on the comparison of reconstruction techniques.
We find that the partial reconstruction of \mttbar\ and \dy\ in \lptj\ sample is fully competitive with that
of the full reconstruction in the \lpgefj\ events.

\begin{table}[htbp]
\caption{Performance of the various reconstruction algorithms for all selected events.
The ranges listed for the \lpgefj\ samples are between the two FOMs derived to ensure
a fair comparison, as detailed in the text.
\label{tab:perf}
}
\begin{center}
\begin{tabular}{llcccc}
&  & \head{\mttbar} & \head{\dy} & \head{$y_l$} & \head{$y_h$}\\
\hline \tablestrut 
& offset & $0.1$ & $0$ & $0.2$ & $0.2$ \\ 
\hline \tablestrut 
 & \multihead{4}{Separation power in \Nsigma}\\
\hline \tablestrut 
\multirow{4}{*}{\lptj} & \chisq\ based & $2.52$ & $2.10$ & $2.00$ & $2.70$ \\
& complete likelihood & $2.53$ & $2.21$ & $2.05$ & $2.69$ \\
&        scaled proxy & $2.60$ & $2.21$ & $2.05$ & $2.69$ \\
&            averaged & $2.65$ & $2.61$ & $2.26$ & $2.92$ \\
\hline
\multirow{2}{*}{\lpgefj}& best assignment & $2.43$--$2.45$ & $1.66$--$1.68$ & $1.57$ & $2.26$ \\
& averaged & $2.53$--$2.56$ & $2.46$--$2.51$ & $1.85$--$1.86$ & $2.70$--$2.71$ \\
\hline
\end{tabular}
\end{center}
\end{table}
%

The \lptj\ channel has the obvious disadvantage of missing a jet.
On the other hand, it has the advantage of fewer jets from initial state radiation, 
and for the algorithm outlined here, of fewer unmatchable events.
These advantages compensate quite well for the missing jet.
It may be that the reconstruction of \lpgefj\ can be improved
by considering additional reconstruction hypotheses, in particular, 
events where one jet is lost and a jet from initial state radiation was selected.

%
%
\section{Summary}
\label{sec:summary}
We present an algorithm that partially reconstructs \ttbar\ events in the \lpj\ channel in the 
case when one of the jets is lost, resulting in a \lptj\ topology.
Probabilities for correct and incorrect jet assignment are formed based on 
$b$-tagging discriminants and on all possible mass combinations on the leptonic and hadronic sides. 
The algorithm can be applied to measure the forward-backward asymmetry in \ttbar\ production, 
the invariant mass spectrum of the \ttbar\ system and for a number of other analyses that require
a full reconstruction.
The performance of the partial reconstruction algorithm is competitive with that commonly achieved 
for fully reconstructed \lpgefj\ events.
The inclusion of \lptj\ events can improve the statistical strength and reduce the
systematic uncertainties of a top properties measurement. 
Gains equivalent to having 50\% more data were achieved at the Tevatron~\cite{bib:afb}.

%
%
%
\section*{Acknowledgments}
We thank our \DZ\ colleagues for useful discussions and for their kind permission to use the \DZ\ 
detector simulation and other collaborative software to expedite the preparation of this \this.
The authors acknowledge the support from the Department of Energy under the grant DE-SC0008475.

\end{document}